\documentclass[a4paper,11pt]{article}
\pdfoutput=1
\usepackage{amsmath,amssymb}
\usepackage{jheppub}
\usepackage{hyperref}
\usepackage{braket}
\usepackage{graphicx}
\usepackage{mathrsfs}
\usepackage{float}
\usepackage{color}%{showkeys}
\newcommand{\beq}{\begin{equation}}
\newcommand{\eeq}{\end{equation}}
\newcommand{\bea}{\begin{equation}\begin{aligned}}
\newcommand{\eea}{\end{aligned}\end{equation}}
\usepackage{myoptions}

\title{Chaos in the butterfly cone}

                                          \author[a]{M\'ark Mezei,}                                       
                                           \author[b,c]{and G\'abor S\'arosi}
                                           \affiliation[a]{Simons Center for Geometry and Physics, SUNY, Stony Brook, NY 11794, USA}                              
                                           \affiliation[b]{David Rittenhouse Laboratory, University of Pennsylvania,\\
                                           Philadelphia, PA 19104, USA}
                                           \affiliation[c]{Theoretische Natuurkunde, Vrije Universiteit Brussels and \\ International Solvay Institutes,\\
Pleinlaan 2, Brussels, B-1050, Belgium}
                                           \emailAdd{mmezei@scgp.stonybrook.edu}                                           
                                           \emailAdd{sarosi@sas.upenn.edu}
                                           
\abstract{
A simple probe of chaos and operator growth in many-body quantum systems is the out of time ordered four point function. In a large class of local systems, the effects of chaos in this correlator build up exponentially fast inside the so called butterfly cone. It has been previously observed that the growth of these effects is organized along rays and can be characterized by a velocity dependent Lyapunov exponent, $\lambda({\bf v})$. 
We show that this exponent is bounded inside the butterfly cone as $\lambda({\bf v})\leq 2\pi T(1-|{\bf v}|/v_B)$, where $T$ is the temperature and $v_B$ is the butterfly speed. This result generalizes the chaos bound of Maldacena, Shenker and Stanford. %In other words, chaos cannot spread faster than the ballistic spreading of holographic systems.
We study $\lambda({\bf v})$ in some examples such as two dimensional SYK models and holographic gauge theories, and observe that in these systems the bound gets saturated at some critical velocity $v_*<v_B$. In this sense, boosting a system enhances chaos. We discuss the connection to conformal Regge theory, where $\lambda({\bf v})$ is related to the spin of the leading large $N$ Regge trajectory, and controls the four point function in an interpolating regime between the Regge and the light cone limit. Finally, we comment on the generalization of the chaos bound to boosted and rotating ensembles and clarify some recent results on this in the literature.
}

\begin{document}
\maketitle

\section{Introduction and summary of results}

An important probe of quantum chaos is the square of the commutator of simple, local operators at different times and positions
\beq
C(t,{\bf x}) = \langle [V(0,{\bf 0}),W(t,{\bf x})]^2 \rangle,
\eeq
where the expectation value is taken in an equilibrium ensemble. In local systems, $V(0,{\bf 0})$ and $W(0,{\bf x})$ commute initially, and $C(t,{\bf x})$ detects the growth of the Heisenberg operator $W(t,{\bf x})=e^{i H t} W(0,{\bf x}) e^{-iH t}$ under  time evolution. In systems with a classical limit, $C(t,{\bf x})$ turns into a Poisson bracket that measures how sensitively the classical trajectory depends on initial data, and hence characterizes the butterfly effect \cite{larkin1969quasiclassical}.

Upon expanding the commutator, one finds that $C(t,{\bf x})$ is given by a combination of four point functions, both time ordered and out of time order (OTO). The former ones typically equilibrate to a constant value in a few local thermalization times, so the latter ones carry the sensitivity to chaos, in particular, they decay to zero after a characteristic scrambling time $t_\text{scr}$ in chaotic systems. In a quantum field theory, it is useful to regulate these correlation functions by distributing operators along the thermal circle. Following \cite{Maldacena:2015waa}, we pick an even distribution and write 
\beq
\label{eq:OTOCdef}
F(t,{\bf x}) = \text{Tr}\big( y V(0,{\bf 0})yW(t,{\bf x}) yV(0,{\bf 0}) y W(t,{\bf x}) \big), \quad y=\frac{e^{-\frac{\beta}{4} H}}{Z^{1/4}},
\eeq
where $Z=\text{Tr}e^{-\beta H}$ and $\beta=T^{-1}$ is the inverse temperature. In systems with a large hierarchy between thermalization and scrambling (i.e. $\beta \ll t_\text{scr}$), this function is expected to have a ``Lyapunov regime", where it behaves like
\beq
\label{eq:OTOCdecay}
F = F_d\left( 1 - \epsilon \,e^{\lambda_L t} + \cdots \right), \quad F_d= \text{Tr}\big( y^2 V(0,{\bf 0})y^2 V(0,{\bf 0}) \big)\text{Tr}\big( y^2 W(0,{\bf 0})y^2 W(0,{\bf 0}) \big),
\eeq
with $F_d$ a time independent constant equal to the disconnected correlator, $\epsilon>0$ a small parameter controlling the separation of time scales, and $ \cdots$ standing for higher powers of $\epsilon\, e^{\lambda_L t}$. This form is valid at intermediate times, when $t$ is past the local thermalization time, but $t\ll t_\text{scr} \sim \log \epsilon^{-1}$. The exponent $\lambda_L$ is the Lyapunov exponent, since for these times, the commutator square behaves as $C(t)\sim \epsilon\, e^{\lambda_L t}$. 

This functional form is very general, in fact it does not assume anything about locality in the theory. For systems with local interactions, one expects some extra structure depending on the spatial separation ${\bf x}$ of the operators. In particular, there will be a butterfly cone, outside of which the commutator $C(t,{\bf x})$ is not growing. For example, in relativistic theories, $C(t,{\bf x})$ must exactly be zero when ${\bf x}^2 > t^2$. The spacetime dependence of $C(t,{\bf x})$ tells us about how the Heisenberg operator spreads, which can happen in various different ways, such as ballistic and diffusive spreading. So one expects that a more refined  understanding of \eqref{eq:OTOCdecay} should be possible for theories with local interactions. 

It was suggested in \cite{Xu:2018xfz,Khemani:2018sdn}, that the growth of $C(t,{\bf x})$ is in general organized along rays ${\bf x} = {\bf v} t$, that is, the spacetime structure is characterized by a \textit{velocity dependent} Lyapunov exponent (VDLE), $\lambda({\bf v})$  \cite{Khemani:2018sdn}:
\beq
C(t,{\bf x}={\bf v}t) \sim e^{\lambda({\bf v}) t}.
\eeq
Therefore, it seems natural to propose
\beq
\label{eq:OTOCVDLE}
f(t,{\bf x}) =1 - \epsilon \,e^{\lambda\left( \frac{{\bf x}}{t} \right)\, t} + \cdots ,\quad \text{where}\quad f(t,{\bf x})=\frac{F(t,{\bf x})}{F_d},
\eeq
as the refinement of \eqref{eq:OTOCdecay} for theories with local interactions.\footnote{Refs.~\cite{Xu:2018xfz,Khemani:2018sdn} mainly focused on this quantity in systems without a hierarchy between thermalization and scrambling, and argued that $\lambda({\bf v})$ is well defined just outside the butterfly cone even in such systems. Here we would like to explore this quantity inside the butterfly cone in systems where the hierarchy between thermalization and scrambling is present.}\footnote{We note that $\ep$ itself can have a weaker than exponential dependence on $t$ and $x$, e.g. a power. $\lam({\bf v})$ contains information about the exponential dependence, and the bounds that we derive are insensitive to the functional form of  $\ep(t,x)$.} Note that when $|{\bf x}|$ is held fixed and $t$ is taken to be large, we recover \eqref{eq:OTOCdecay} with $\lambda_L=\lambda({\bf 0})$.
We can also define the butterfly cone in terms of the VDLE: it is the set of velocities where $\lambda({\bf v})$ vanishes, in particular, for isotropic systems, $\lambda(v_B {\bf \hat n})=0$, where ${\bf  \hat n}$ is any unit vector and $v_B$ is the butterfly velocity.

Under some rather general assumptions that we will review, Maldacena, Shenker and Stanford (MSS) \cite{Maldacena:2015waa} showed that 
\beq
\label{eq:MSS}
\frac{|\partial_t f(t,{\bf x})|}{1-f} \leq \frac{2\pi}{\beta}.
\eeq
For the non-local ansatz \eqref{eq:OTOCdecay}, this implies the bound $\lambda_L \leq \frac{2\pi}{\beta}$. 
For the local ansatz \eqref{eq:OTOCVDLE}, we get instead the bound 
\beq
\label{eq:VDLObound1}
|\lambda({\bf v})-{\bf v}\cdot \nabla \lambda({\bf v})|\leq \frac{2\pi}{\beta},
\eeq
namely, the \textit{Legendre transform} of the VDLE is bounded for all velocities. In the rotational invariant case the VDLE depends only on $v=|{\bf v}|$ and the
$\lambda(v)$ that saturates the bound solves a first order ODE. The solution is $\lambda(v)=\frac{2\pi}{\beta}\le(1-\frac{v}{v_B}\ri)$, where $v_B$ is so far an integration constant. From the definition of the butterfly velocity $\lambda(v_B)=0$, we learn that the integration constant is also the butterfly velocity. This is the ballistic butterfly front of a holographic system.\footnote{For holographic systems, $v_B$ has a particular value that is not constrained by the present discussion.} Using a simple argument, we will show that \eqref{eq:VDLObound1} in fact implies the universal bound 
\beq
\label{eq:VDLObound2}
\lambda(v)\leq \frac{2\pi}{\beta}\left(1-\frac{v}{v_B}\right).
\eeq
It is also simple to generalize this bound for asymmetric butterfly cones;	we will discuss this in the main text. 

\begin{figure}[!h]
\begin{center}
\includegraphics[scale=0.6]{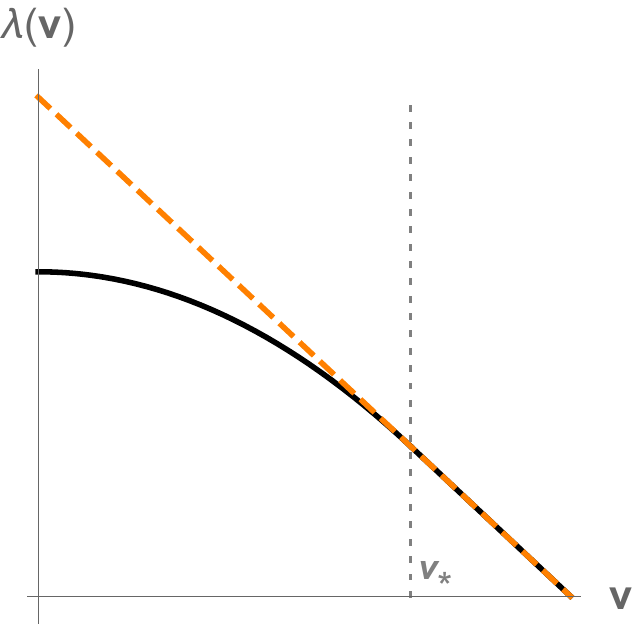}
\caption{Illustration of a typical velocity dependent Lyapunov exponent in strongly coupled theories. At small velocities, the function is quadratic, which corresponds to a diffusive spreading of operators. The VDLE saturates the bound \eqref{eq:VDLObound2} from some finite, order one critical velocity $v_*$, giving a ballistic butterfly front with maximal Lyapunov exponent. \label{fig:toyvdle}}
\end{center}
\end{figure}

In principle, it is possible to saturate the bound \eqref{eq:VDLObound2} for some subset of velocities, while not for others. In particular, one could have a critical velocity $v_* <v_B$, such that the bound is saturated only for $v>v_*$ (see Fig.~\ref{fig:toyvdle}).  We show that this happens in any chaotic large $N$ 2d CFT.  It is also a rather generic situation for SYK chains \cite{Gu:2016oyy} and other higher dimensional generalizations of the SYK model \cite{Murugan:2017eto,Peng:2018zap,Lian:2019axs}.\footnote{The existence of a critical velocity $v_*$ above which the VDLE is linear in $v$ was also explored by Gu and Kitaev in SYK chains \cite{Gu:2018jsv}, see also \cite{Guo:2019csw}. In this paper, we interpret this regime as having maximal chaos. For the SYK chain, \cite{Guo:2019csw} speculated that the microscopic explanation of this phenomenon could be that the Schwarzian mode spreads faster than the conformal matter. Here we find this behavior rather universal, which raises the question if there is a more universal microscopic explanation as well.} It also happens in holographic gauge theories, when stringy corrections are taken into account \cite{Shenker:2014cwa}. We will review each of these examples and discuss the corresponding VDLE.

The existence of $v_*<v_B$ has a striking signature in the OTOC for fixed $\abs{{\bf x}}\gg \beta$, as was understood in \cite{Gu:2016oyy,Gu:2018jsv}: for times obeying $\abs{{\bf x}}/v_B< t <\abs{{\bf x}}/v_*$ the OTOC exhibits maximal growth in time with $\lam_L=2\pi/\beta$. For very large $\abs{{\bf x}}$ we also have to take into account the condition that the OTOC stays in the Lyapunov regime, that is, it has not yet started to saturate due to higher order effects in $\epsilon e^{\lambda({\bf x}/t)t}$. This defines the edge of the scrambling region, $t_{\rm scr}({\bf x})$, via solving
\beq
\lambda \left( \frac{{\bf x}}{t_{\rm scr}({\bf x})} \right)t_{\rm scr}({\bf x}) \sim \log \epsilon^{-1}.
\eeq
The VDLE bound \eqref{eq:VDLObound2} then implies that $t_{\rm scr}({\bf x})\geq \frac{\beta}{2\pi}\log \epsilon^{-1} + |\bf{x}|/v_B$, which is a local version of the fast scrambling bound. Notice that in the case of non-maximal chaos, when the VDLE has a shape like in Fig. \ref{fig:toyvdle}, the edge of the scrambling region is no longer a cone, but its tip gets smoothed out \cite{Shenker:2014cwa}.

 %condition $t\ll \abs{{\bf x}}/v_B + t_\text{scr}$.

When the theory in question is a CFT on hyperbolic space, with the radius of curvature equal the temperature, the OTO four point functions are related to certain analytic continuations of the flat space four point function. (This is also the case for thermal 2d CFTs on the line.)  To define the VDLE for this situation, we replace $|{\bf x}|$ in \eqref{eq:OTOCVDLE} with the geodesic distance between the operators on hyperbolic space. In this case, we will show that the VDLE characterises a one parameter family of limits interpolating between the Regge ($v=0$) and light cone ($v=1$) limits. We will find that the VDLE is directly related to the spin of the leading large $N$ Regge trajectory, while the critical velocity $v_*$ is related to the derivative of this trajectory at the stress tensor, which in turn is constrained by the presence of non-conserved higher spin operators. We will see that above this critical velocity $v_*$, the limit is dominated by the stress tensor in the light cone OPE, below $v_*$ it is dominated by the pomeron (leading Regge trajectory). This happens via an exchange of dominance between a saddle point and a kinematic pole associated to the stress tensor in the integral that results from resummation of the OPE in the Regge limit. This is very reminiscent to how the critical velocity $v_*$ comes about in SYK-like models; in those cases, the presence of the pole is enforced by the ladder identity of Gu and Kitaev \cite{Gu:2018jsv}. In fact, these two mechanisms are the same for the models of Murugan, Stanford and Witten (MSW) \cite{Murugan:2017eto} which flow to 2d CFTs in the IR. Another corollary of the Regge analysis will be a bound on the Rindler space butterfly speed $v_B\leq (d-1)^{-1}$, valid for any large $N$ CFT$_d$.

In holographic gauge theories $v_*$ is proportional to the inverse 't Hooft coupling. For SYK-like CFTs $v_*$ is just some order one number.
It is interesting to ask how $v_*$ changes if we move in a family of CFTs. On one hand, in $d=2$ we show that $v_*\to v_B=1$ as the theory becomes weakly coupled. On the other hand, for $d>2$ we argue that, at least on hyperbolic space, $v_*=v_B$ for some finite coupling, and hence the region of maximal chaos is not visible in perturbation theory. The argument makes use of the assumed convexity of the leading large $N$ Regge trajectory to put a lower bound on $v_*$ in terms of the Rindler space Lyapunov exponent. The bound is such that it forces $v_*\to v_B$ as $\lam_L\to 0$ in $d=2$, while in $d>2$ must reach $v_*\to v_B$ before the Lyapunov exponent would reach zero.

OTOCs were computed in thermal weakly coupled systems by summing classes of Feynman diagrams \cite{Stanford:2015owe,Aleiner:2016eni,Chowdhury:2017jzb,Steinberg:2019uqb}.\footnote{See also \cite{deMelloKoch:2019ywq} for the determination of $\lam_L$ for all values of the coupling in the nonunitary fishnet theories. Our discussion only applies to unitary theories.} Our bound \eqref{eq:VDLObound2} applies to these results, and it would be interesting to extract the VDLE from the formulas in the literature. This would require precision numerics.  For relativistic theories it was found that $v_B\approx 1$ \cite{Chowdhury:2017jzb,Steinberg:2019uqb}, and it would be interesting to understand, if there exists a $v_*<v_B$ for these systems.

Another interesting question is what we can  say about the development of chaos in  ensembles other than the thermal one. In case we only turn on chemical potentials for spacetime charges, there is a fairly obvious version of the MSS bound that still applies to the rate of change in the direction in which the combination of charges in the ensemble translates. A much less trivial question is what we can  say about the time derivative alone in such ensembles. The simplest example is when the ensemble in question is a boosted thermal ensemble in a Lorentz invariant theory. In this case, the rate of growth in the temporal direction is just determined by the same velocity dependent Lyapunov exponent as defined in \eqref{eq:OTOCVDLE}, where ${\bf v}$ is the amount of boost. For example, in two dimensional CFTs, we have $v_B=1$, and the bound \eqref{eq:VDLObound2} is then equivalent with
\beq
\label{eq:chiraltemperaturebound}
\lambda_L \leq \min \left\lbrace \frac{2\pi}{\beta_+}, \frac{2\pi}{\beta_-} \right\rbrace,
\eeq
in a boosted ensemble with left and right moving inverse temperatures $\beta_-$ and $\beta_+$.\footnote{For the boosted BTZ black brane, the scrambling time, as defined via the mutual information, is controled by the smaller chiral temperature \cite{Stikonas:2018ane}, consistently with our bound. } 
 
 Another relevant example is when we put the theory on a sphere and turn on an angular velocity; the proof of MSS \cite{Maldacena:2015waa} does not apply to the temporal derivative in this setup.\footnote{A naive generalization of the MSS proof technique yields a bound on a combination of temporal and angular derivatives, see sec. \ref{sec:Rotating}. Trying to apply the method to the temporal derivative alone, it turns out that the Cauchy-Schwarz inequality does not provide a tight enough bound on the magnitude of the OTOC on the edges of the strip of analyticity.} The boosted thermal ensemble can be thought of as an infinite volume limit of this, and hence the bound \eqref{eq:VDLObound2} should apply at early times. On the other hand, it is difficult to say anything about the growth rate for the finite size rotating system, for example it is not clear if even the bound $\lambda_L \leq 2\pi /\beta$ holds.  One can gain some intuition from holographic calculations. For example, in a holographic 2d CFT, one can calculate the OTOC in a rotating ensemble, by repeating the Shenker-Stanford shockwave calculation \cite{Shenker:2013pqa,Roberts:2014isa,Shenker:2014cwa} for rotating BTZ. This has been done in \cite{Poojary:2018esz,Jahnke:2019gxr}, and these authors concluded that the bound $\lambda_L \leq 2\pi /\beta$ is not obeyed. We will re-examine these calculations and will reach a different conclusion. We will find that the growing part of the OTOC is not purely exponential, but it is modulated by a periodic function, with period determined by the size of the spatial circle. On average, the Lyapunov exponent is $\lambda_L= 2\pi /\beta$. On the other hand, the instantaneous bound \eqref{eq:MSS} can be violated. However, the violation only happens after some time that scales with the size of the system. When the circle size is comparable to the scrambling time, no violation of \eqref{eq:MSS} can be observed, in fact in this case, the system is effectively infinite size in the Lyapunov regime and the stronger bound \eqref{eq:chiraltemperaturebound} holds. Finally, let us stress that since the bound \eqref{eq:chiraltemperaturebound} comes directly from \eqref{eq:VDLObound2}, it applies only to OTOCs of \textit{local} operators and is therefore not in tension with the results of \cite{Reynolds:2016pmi}, who found that $\lambda_L = 2\pi /\beta$ for global shocks in rotating black holes.

This paper is organized as follows. In sec.~\ref{sec:Bounds} we prove a bound on the VDLE $\lam({\bf v})$. In sec.~\ref{sec:Examples} we consider concrete systems where the VDLE has been computed in the literature and examine the interplay of these results with the bounds. In sec.~\ref{sec:Regge} we relate the VDLE of the Rindler OTOC to the spin of the leading large $N$ conformal Regge trajectory and derive constraints on the VDLE from Regge physics. We end with an analysis of chaos in rotating ensembles in sec.~\ref{sec:Rotating}.

%\pagebreak
\section{The general argument}\label{sec:Bounds}

\subsection{Review of the MSS bound}

In \cite{Maldacena:2015waa}, MSS showed that if a function $f(t)$ is
\begin{enumerate}
\item \label{it:1} Analytic in the strip $|\text{Im}t|\leq \beta/4$
\item \label{it:2} Real for real $t$
\item \label{it:3} Bounded by one on the half strip: $|f|\leq 1$ on $\text{Re}\,t>t_0$
\end{enumerate}
then it satisfies
\beq
\frac{|\partial_t f|}{1-f} \leq \frac{2\pi}{\beta} + O(e^{-\frac{4\pi}{\beta}(t-t_0)}).
\eeq
They then applied this bound to the normalized OTO correlation function
\beq
\label{eq:OTOCdef2}
f(t,{\bf x}) = \frac{\text{Tr}\big( y V(0,{\bf 0})yW(t,{\bf x}) yV(0,{\bf 0}) y W(t,{\bf x}) \big)}{\text{Tr}\big( y^2 V(0,{\bf 0})y^2 V(0,{\bf 0}) \big)\text{Tr}\big( y^2 W(0,{\bf 0})y^2 W(0,{\bf 0}) \big)}, \quad y=\frac{e^{-\frac{\beta}{4} H}}{Z^{1/4}}.
\eeq
This function satisfies~\ref{it:1} and~\ref{it:2} when $V$ and $W$ are Hermitian operators and the Hamiltonian is bounded from below. In order to have~\ref{it:3}, one needs to assume that time ordered correlation functions factorize for times larger than the local thermalization time $t_d \sim \beta$, and that OTOCs approximately factorize around a time $t_0$ that is larger than the local thermalization time, but much shorter than the scrambling time, $t_d< t_0 \ll t_s$. These assumptions are automatic in a large $N$ theory, or other theories with classical limits, but they are expected to hold more generally for chaotic theories with many local degrees of freedom. Condition~\ref{it:3} follows from these factorization assumptions by applying the Cauchy-Schwarz inequality (which requires unitarity) and the maximum modulus principle inside the strip.

\subsection{VDLE bound in isotropic systems}

As explained in the Introduction, the MSS bound \eqref{eq:MSS} applied to the formula \eqref{eq:OTOCVDLE} defining the velocity dependent Lyapunov exponent $\lambda({\bf v})$ gives the bound \eqref{eq:VDLObound1} on the Legendre transform of $\lambda({\bf v})$. Let us first examine this bound for systems with rotational invariance. In this case, the VDLE can only depend on the magnitude of the velocity $v=|{\bf v}|$ and \eqref{eq:VDLObound1} reads as
\beq
\label{eq:Legendreboundisotropic}
|\lambda(v)-v \lambda'(v)|\leq \frac{2\pi}{\beta}.
\eeq
We want to derive a bound on $\lam(v)$ starting from \eqref{eq:Legendreboundisotropic}. To this end, following the proof technique of \cite{Mezei:2016zxg}, we rewrite the inequality as a differential equation
\beq
\lambda(v)-v \lambda'(v)= a(v)\,, \qquad | a(v)| \leq \frac{2\pi}{\beta},
\eeq
and solve it subject to the boundary condition $\lam(v_B)=0$. In terms of $a(v)$, the VDLE then reads as
\es{lamsol0}{
\lambda(v)=v\int^{v_B}_v du \frac{a(u)}{u^2}\,, \quad v\leq v_B\,.
}
Using this, we can bound the VDLE directly
\es{VDLEBound}{
|\lambda(v)| &\leq v\int^{v_B}_v du \frac{|a(u)|}{u^2} \\
&\leq \frac{2\pi}{\beta}v\int^{v_B}_v du \frac{1}{u^2}\\
&=\frac{2\pi}{\beta} \left( 1-\frac{v}{v_B} \right),
}
which is the bound announced in the Introduction.

\subsection{VDLE bound in anisotropic systems}

We start our analysis of anisotropic systems with $d=2$.
In $d=2$ the velocity lives on the line, it is just a number with sign, which we will denote here with $v$. Having an asymmetric butterfly cone means that we have left and right butterfly velocities $v_B^{\pm}$ and $v_B^+\neq -v_B^-$. The generalization of \eqref{lamsol0} for this case is:
\es{lamsol}{
\lam(v)&=\begin{cases}
-v\int_{v_B^-}^v du \ {a(u)\ov u^2}\,, \qquad &\text{if $v$ and $v_B^-$ are of the same sign,}\vspace{0.5cm}\\
v\int^{v_B^+}_v du \ {a(u)\ov u^2}\,,\qquad &\text{if $v$ and $v_B^+$ are of the same sign.}
\end{cases}
}
The reason for the restriction on the sign of $v$ is that the integrand in general is not integrable at $0$, and only the appropriate line makes sense.
There are two cases to consider. If $v_B^\pm>0$, we can use both lines to compute $\lam(v)$. (This also holds for $v_B^\pm<0$.) For $v_B^-<0<v_B^+$, we have to use the first line for $v<0$ and the second for $v>0$.  

\begin{figure}[!h]
\begin{center}
\includegraphics[scale=0.5]{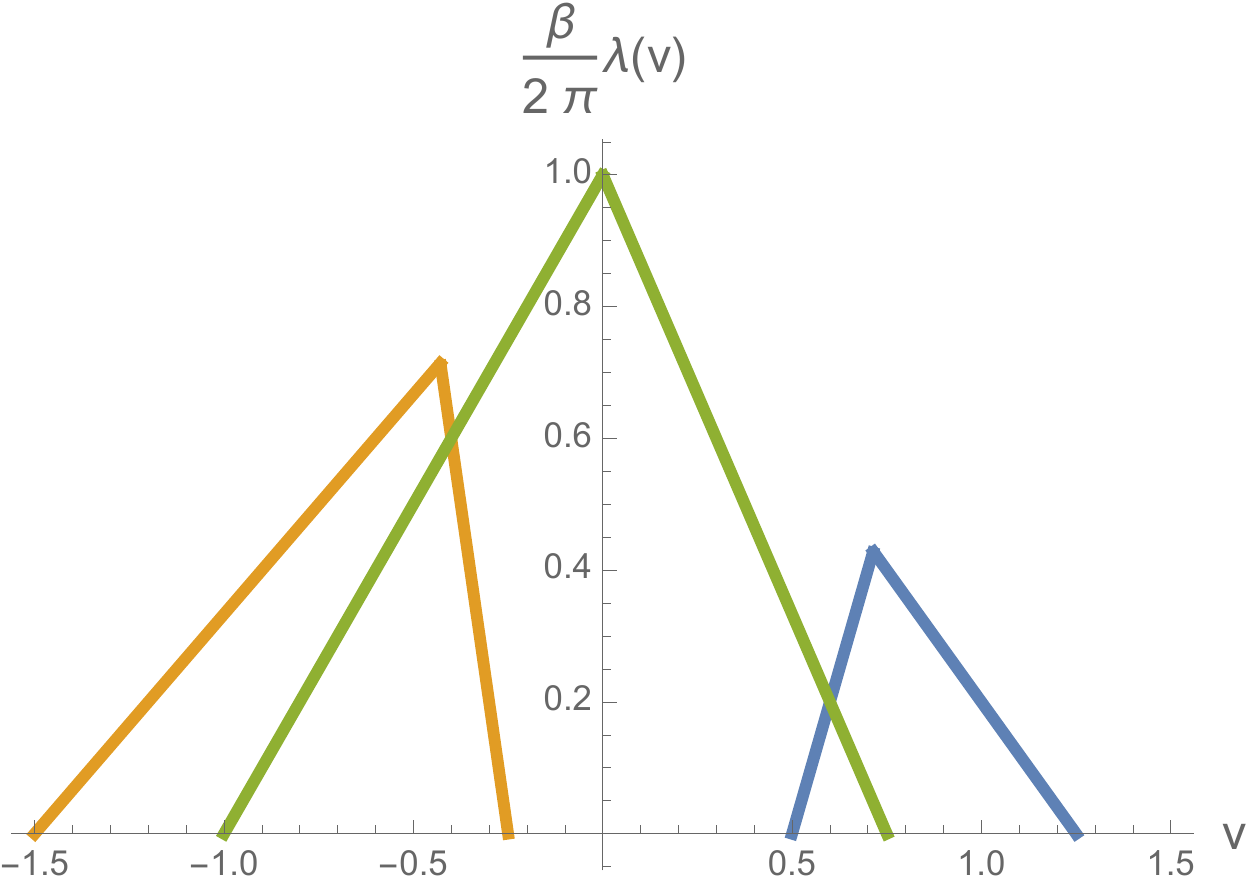}
\caption{The bound  \eqref{BoundSummary} for various values of $v_B^\pm$ and for $\beta=2\pi$.\label{fig:lambdaBound}}
\end{center}
\end{figure}

Bounding the VDLE goes exactly as in \eqref{VDLEBound}. By enumerating all cases,  we get the simple result
\beq
\label{BoundSummary}
\lambda(v)\leq {2\pi\ov \beta} \min\left\lbrace \left| 1-\frac{v}{v_B^-}\right|, \left| 1-\frac{v}{v_B^+}\right|\right\rbrace\,.
\eeq
This holds irrespective of the signs of $v,\, v_B^{\pm}$.
 We plot three cases of \eqref{BoundSummary} on Fig.~\ref{fig:lambdaBound}. We will apply this bound to a model with an asymmetric butterfly cone in sec.~\ref{sec:chiralSYK}.

\begin{figure}[!h]
\begin{center}
\includegraphics[scale=0.4]{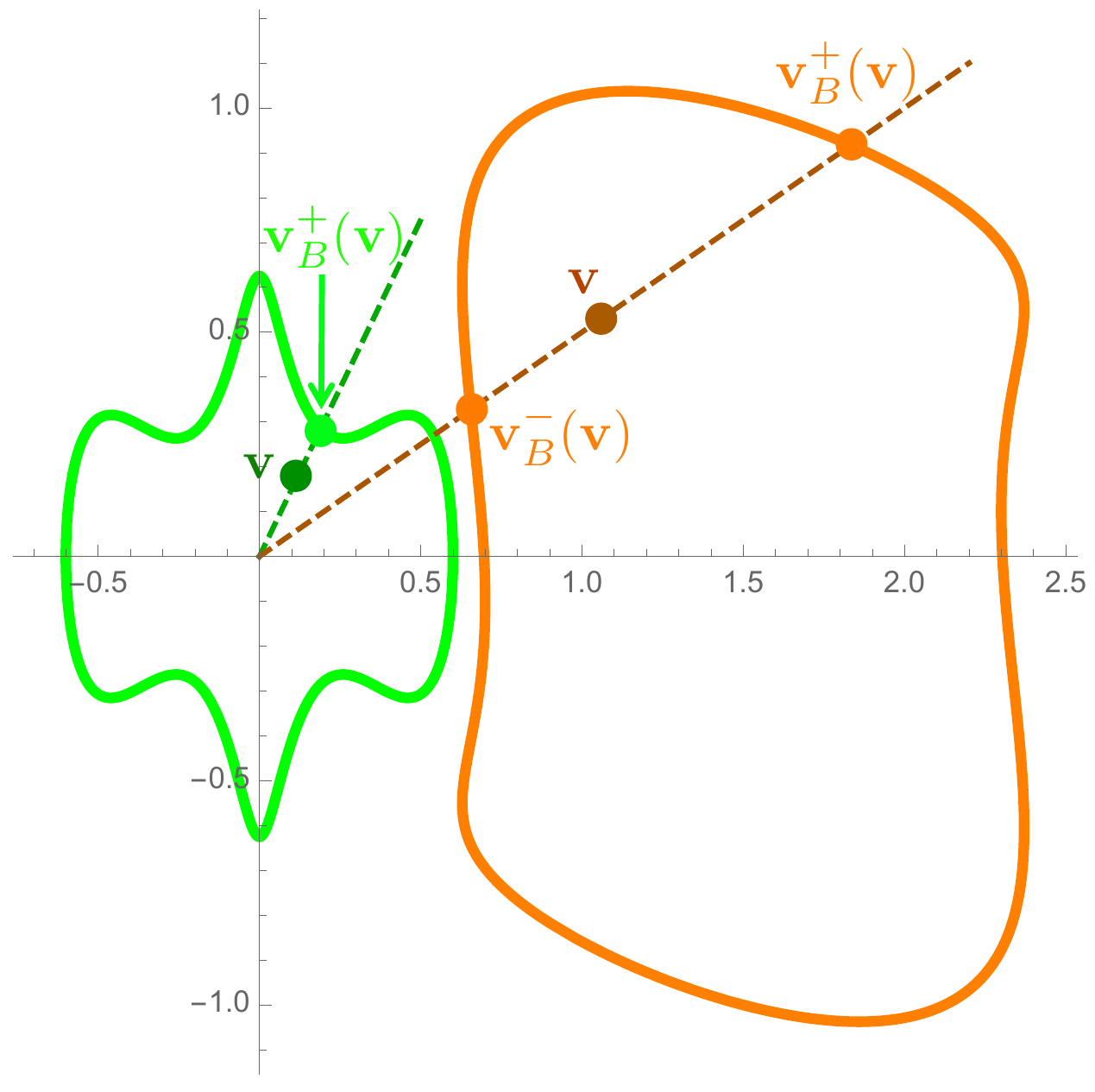}
\caption{Definition of ${\bf v}_B^{\pm}({\bf v})$. The dashed half lines connect ${\bf v}$ with the origin, and their intersection with the shapes define ${\bf v}_B^{\pm}({\bf v})$. For the green shape only ${\bf v}_B^+({\bf v})$ is defined. For the orange shape ${\bf v}_B^{+}({\bf v})$ is the farther and  ${\bf v}_B^{-}({\bf v})$ is the closer intersection.    \label{fig:vBpmDef}}
\end{center}
\end{figure}

It is straightforward to generalize this idea for asymmetric butterfly cones in higher dimensions. Let us define the butterfly cone in terms of a subset of velocities $\mathcal{C}_B$ as
\beq
\mathcal{C}_B = \lbrace {\bf v}| \lambda({\bf v})=0 \rbrace.
\eeq
Let us choose a velocity ${\bf v}$ and take the half line from the origin through ${\bf v}$. It intersects $\mathcal{C}_B$ in ${\bf v}_B^+({\bf v})$ and possibly in ${\bf v}_B^-({\bf v})$ as shown in  Fig.~\ref{fig:vBpmDef}.

\begin{figure}[!h]
\begin{center}
\includegraphics[scale=0.7]{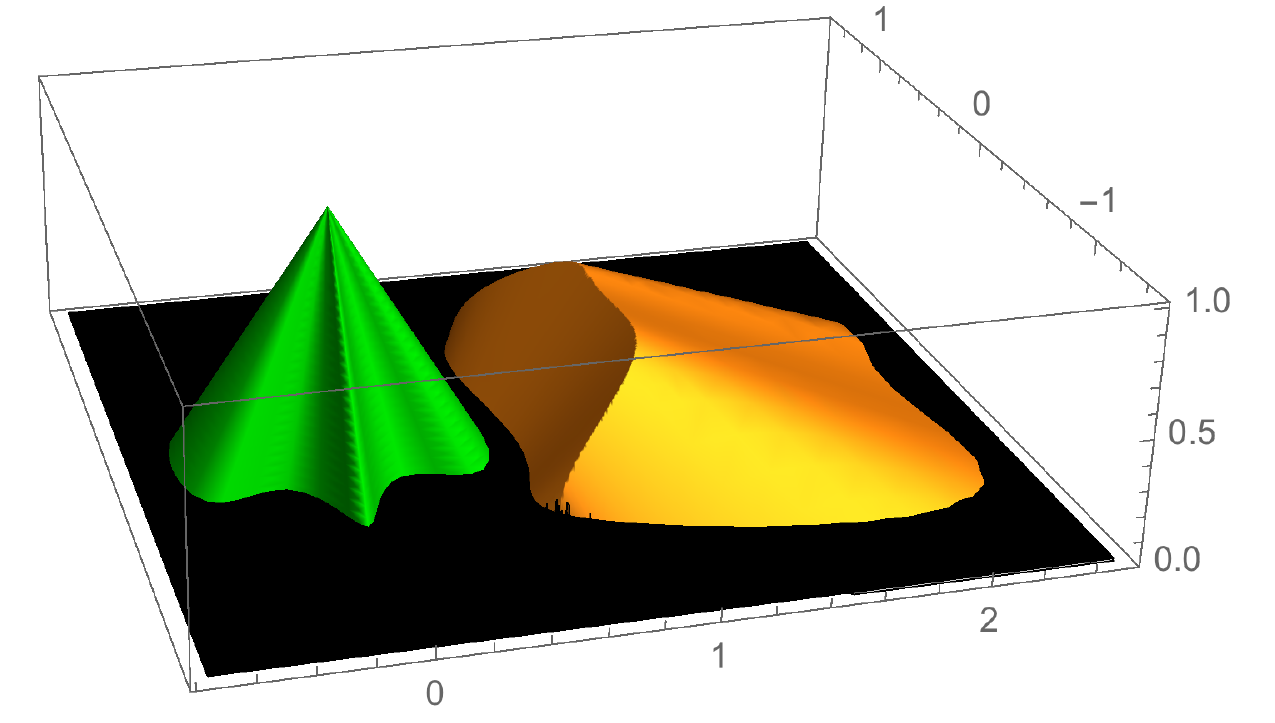}
\caption{The upper bound \eqref{BoundSummary2} for $\lam({\bf v})$ for the two shapes considered in  Fig.~\ref{fig:vBpmDef} (and for $\beta=2\pi$).   \label{fig:anisBound}}
\end{center}
\end{figure}

We rewrite \eqref{eq:VDLObound1} as
\beq
\lambda({\bf v})-{\bf v}\cdot \nabla \lambda({\bf v})= a({\bf v})\,, \qquad | a({\bf v})| \leq \frac{2\pi}{\beta}\,,
\eeq
and  can write $ \lambda({\bf v})$ in terms of $a({\bf v})$ as a line integral
\beq
\lambda({\bf v}) = \int^{{\bf v}_B^\pm({\bf v})}_{{\bf v}}( d{\bf u} \cdot {\bf v})\frac{a({\bf u})}{|{\bf u}|^2}, 
\eeq
where the path is taken to be the half line from Fig.~\ref{fig:vBpmDef}. Since ${\bf u}$ is parallel to ${\bf v}$ these integrals are of the form \eqref{lamsol}, and can be bounded as in \eqref{BoundSummary}. The final result is
\beq
\label{BoundSummary2}
\lambda({\bf v})\leq {2\pi\ov \beta} \min\left\lbrace 1-\frac{\abs{{\bf v}}}{\abs{{\bf v}_B^+({\bf v} )}},\frac{\abs{{\bf v}}}{\abs{{\bf v}_B^-({\bf v} )}}-1\right\rbrace\,,
\eeq
where we just take the first argument of the minimization, if ${\bf v}_B^-({\bf v} )$ does not exist. Two examples are shown on Fig.~\ref{fig:anisBound} to help understand the bound. 

We may wonder if we have found an ideal bound. Let us take
\es{lamMaxAnis}{
\lambda_\text{max}({\bf v})\equiv {2\pi\ov \beta} \min\left\lbrace 1-\frac{\abs{{\bf v}}}{\abs{{\bf v}_B^+({\bf v} )}},\frac{\abs{{\bf v}}}{\abs{{\bf v}_B^-({\bf v} )}}-1\right\rbrace\,.
}
We can verify that $\lambda_\text{max}({\bf v})$ saturates the bound \eqref{eq:VDLObound1} for all ${\bf v}$ by explicit computation. All we have to use is that (locally) ${\bf v}_B^\pm({\bf v} )$ only depends on the direction of ${\bf v}$ that we denote by ${\bf \hat v}$, and that ${\bf v}_i  \p_i \,{\bf \hat v}_j=0$. Thus, we conclude that we have found an ideal bound in \eqref{BoundSummary2}.

\section{Examples}\label{sec:Examples}

\subsection{SYK chain}

The SYK model \cite{Sachdev:1992fk,Polchinski:2016xgd,Maldacena:2016hyu} is a model of interacting Majorana fermions with random all-to-all couplings. It drew considerable attention in recent years as an example of a strongly coupled system that is solvable at large $N$. An interesting feature of this model is that it displays a universal pattern of conformal symmetry breaking in the IR that is shared with near extremal black holes that have a long, nearly AdS$_2$ throat, and therefore it can be viewed as a toy model for these black holes \cite{Sachdev:2015efa,Maldacena:2016upp}, see \cite{Sarosi:2017ykf,Rosenhaus:2018dtp} for reviews. The OTO correlator displays a maximal Lyapunov exponent $\lambda_L=2\pi \beta^{-1}$ to leading order in the inverse coupling. However, the model has no spatial locality, so it is not adequate to explore the velocity dependent Lyapunov exponent. Fortunately, many higher dimensional generalizations of the model have been constructed, which display similar physics with additional spatial locality. 

\begin{figure}[!h]
\begin{center}
\includegraphics[scale=0.3]{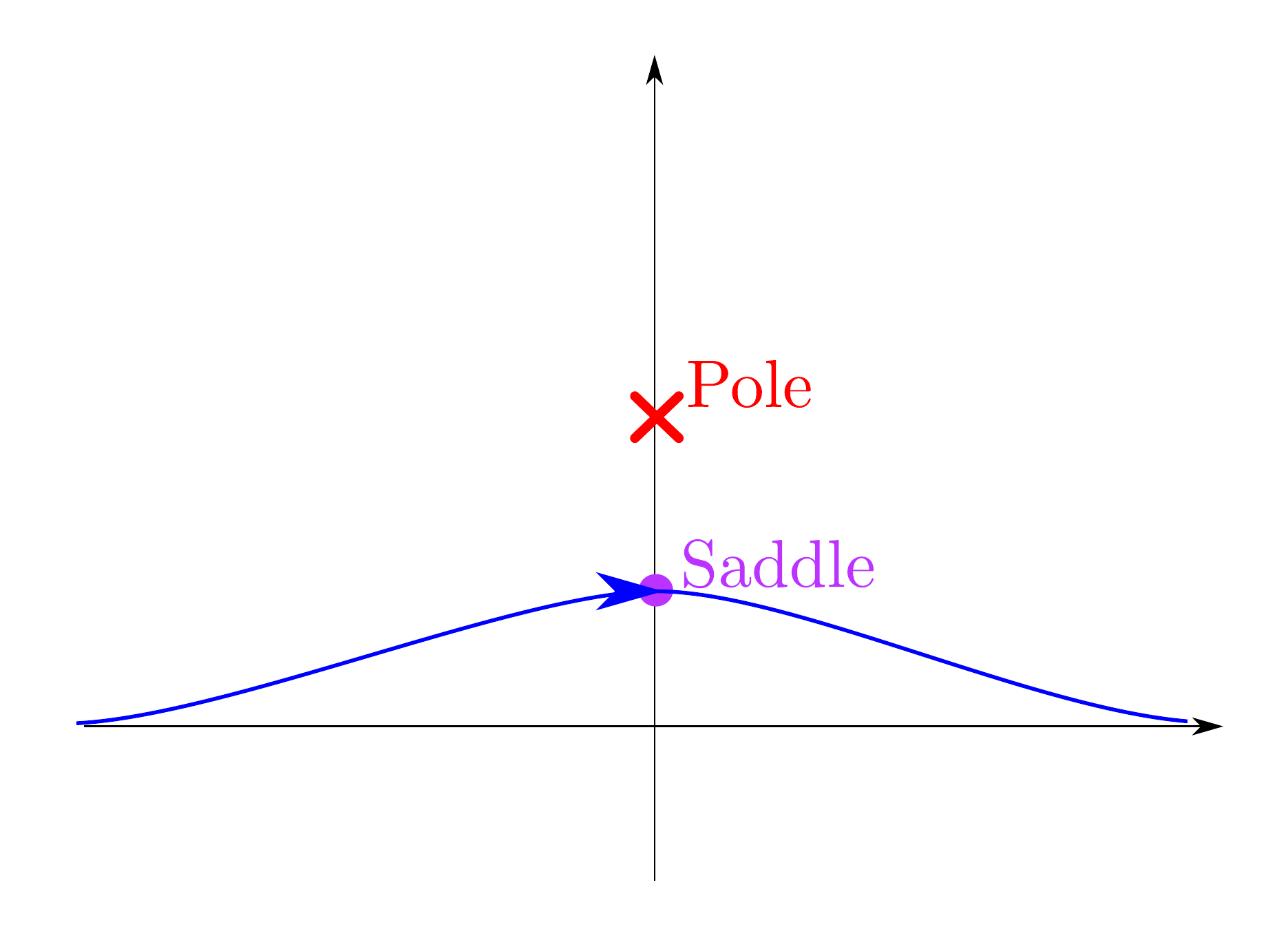}\includegraphics[scale=0.3]{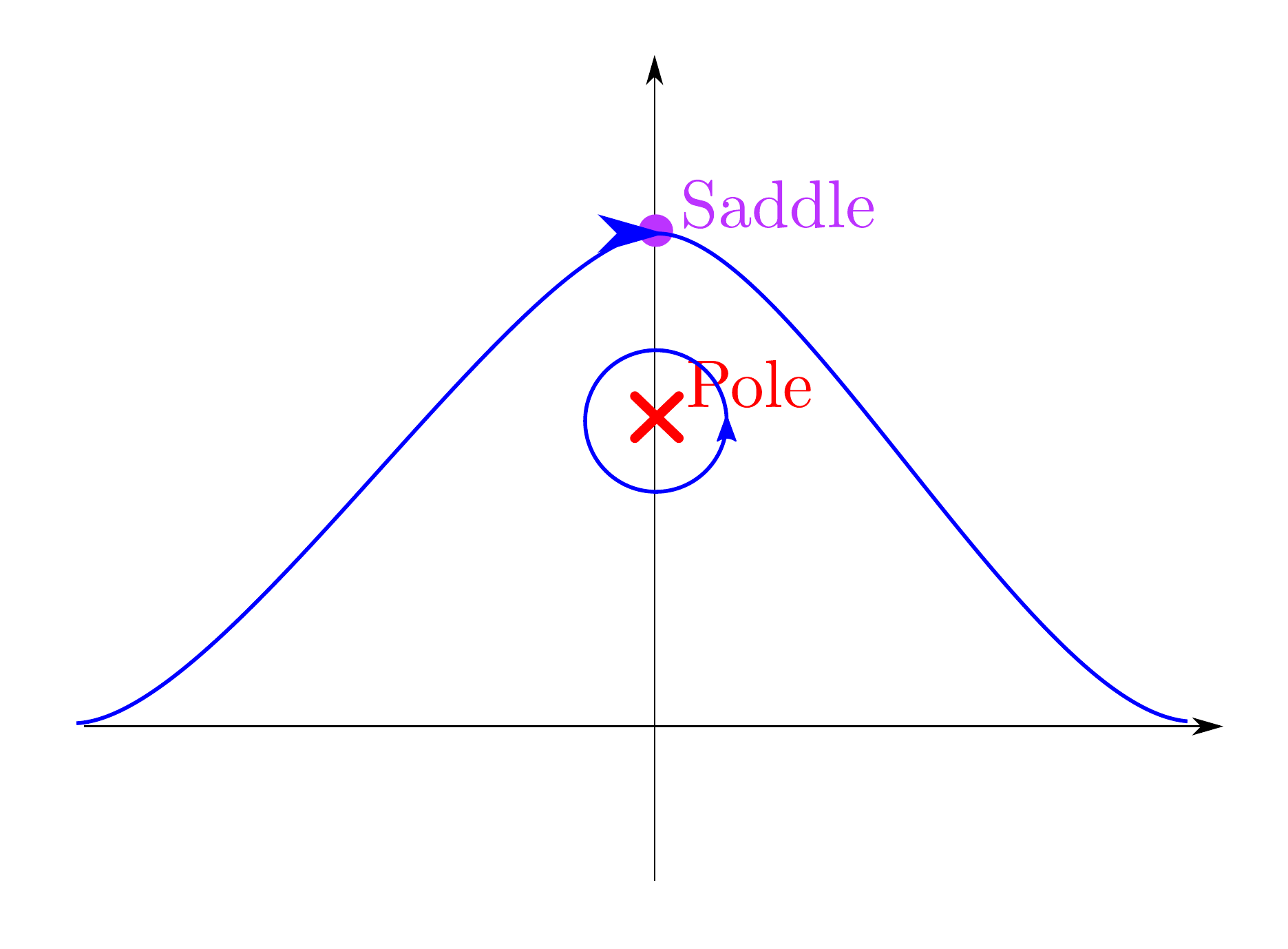}
\caption{The generic mechanism for saturating the bounds \eqref{eq:VDLObound1}-\eqref{eq:VDLObound2} above some critical velocity. The growing contribution to the OTOC is given by an integral, that can be evaluated by saddle point for large $t$. The integration contour is drawn by a blue solid line. As we increase the velocity $v$, the saddle point shifts in the direction of the imaginary axis, and eventually it crosses a pole. After this, the contribution of the pole dominates the integral. \label{fig:exchangeofdom}}
\end{center}
\end{figure}

The simplest such model is obtained by considering $M$ copies of the SYK model in a one dimensional chain, and introducing local (four body) couplings between them \cite{Gu:2016oyy}.\footnote{One can also introduce instead two body couplings between the sites \cite{Song:2017pfw}.} 
This model has two parameters; $J_0$ is the width of the distribution of the couplings of each individual SYK model, and $J_1$ is the width of the distribution of the couplings between different SYK sites. The effective coupling is $J=\sqrt{J_0^2+J_1^2}$. The OTOC \eqref{eq:OTOCdef} of the fermion operators for $\beta J\gg 1$ takes the form \cite{Gu:2016oyy}
\es{SYKChainInt}{
f(t,x) \approx 1-\frac{1}{N} \int_{-\infty}^\infty \frac{dp}{2\pi} \frac{e^{ipx}}{b(p)} e^{\frac{2\pi}{\beta}[1-3b(p)]t}, \quad b(p)=\frac{\alpha}{J} \left( \frac{2\pi}{\beta}+D p^2 \right),
}
where $\alpha$ is some order one number, and $D\sim J_1^2J^{-1}$ is the diffusion constant of energy transport in the model. As explained in \cite{Gu:2016oyy}, at large $t$ there are two possible behaviors for this integral. When $x$ is small, the integral is dominated by a saddle point. When $x$ is large, it is dominated by the pole at $b(p)=0$. 
This can be made particularly transparent using the definition \eqref{eq:OTOCVDLE} of the velocity dependent Lyapunov exponent. We substitute $x=vt$ and take $t$ large. We can then evaluate the integral by saddle point, but increasing $v$ shifts the saddle in the imaginary $p$ direction. Eventually, at $v=v_*$ the saddle crosses the pole coming from $b(p)$ in the denominator, so to deform into the contour of steepest descent, we need to pick up the contribution of the pole, see Fig.~\ref{fig:exchangeofdom}. After this, the pole actually gives a larger contribution than the saddle. 
This mechanism was understood before us; very similar discussions can be found in \cite{Xu:2018xfz,Gu:2018jsv}.

\begin{figure}[!h]
\begin{center}
\includegraphics[width=0.4\textwidth]{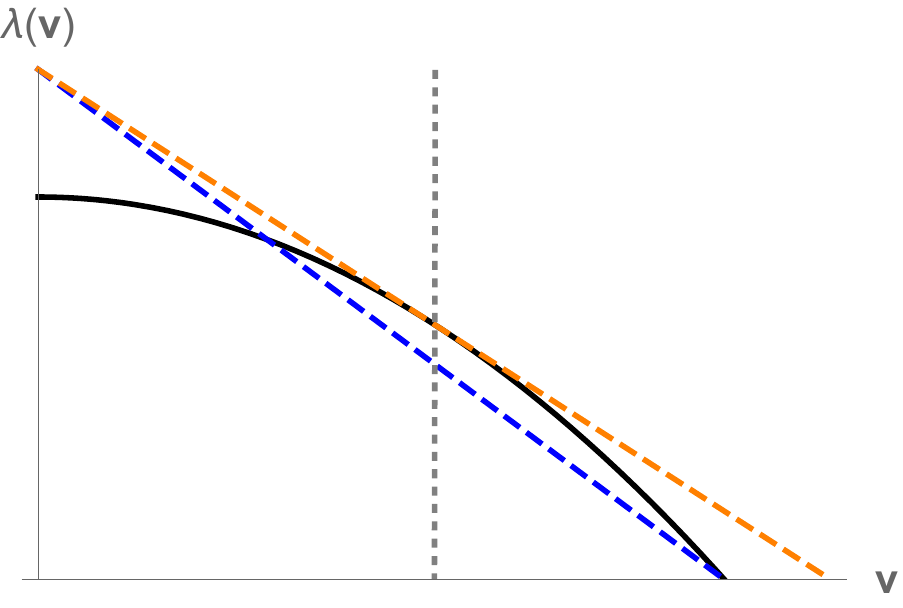}
\caption{The black curve is a typical saddle point contribution to the VDLE. If it was the dominant contribution to the VDLE for all $v$, it would violate the bound \eqref{eq:VDLObound2} drawn in dashed blue. To obey the bound \eqref{eq:VDLObound2}, we need to increase the butterfly velocity until the bound touches the black curve (this is the orange dashed line). The critical velocity $v_*$ is where the two curves touch (gray, dotted), and for $v>v_*$, the saddle cannot dominate.  \label{fig:boundvssaddle}}
\end{center}
\end{figure}

As we will see, this mechanism turns out to be very generic; it applies to all SYK-like models, to holographic gauge theories, to generic large $N$ thermal 2d CFTs on the line, and to higher dimensional large $N$ thermal CFTs on hyperbolic space. A novelty in our presentation is that we attribute a new property to the regime $v\geq v_*$: the VDLE saturates the bound \eqref{eq:VDLObound2} (or its anisotropic version), when the pole contribution dominates the OTOC. We can in fact think about the bound as enforcing the presence of the pole: in all the examples with $v_*<v_B$ that we are going to discuss, the bound \eqref{eq:VDLObound2} would have been violated by the saddle point contribution alone. The critical velocity is the velocity beyond which the saddle cannot give the dominant contribution such that the VDLE still obeys \eqref{eq:VDLObound2}, therefore a pole must take over. This is illustrated in Fig.~\ref{fig:boundvssaddle}.

%\pagebreak
The critical velocity in the SYK chain can be determined by equating the location of the saddle point and the pole, giving
\es{vsvbSYK}{
v_*=\frac{12\pi \alpha v_B}{\beta J}, \quad v_B^2={2\pi D\ov  \beta}
}
where $v_B$ is the butterfly velocity. Note that since the derivation only applies in the $\beta J\gg 1$ limit, we always have $v_*\ll v_B$.  The velocity dependent Lyapunov exponent is
\beq
\label{eq:VDLEchain}
\lambda(v)=
\begin{cases}
 \frac{\pi}{\beta} \left( 2-\frac{v_*}{v_B}-\frac{v^2}{v_* v_B}\right) , & v<v_*\,,\\
 \frac{2\pi}{\beta}\left(1-\frac{v}{v_B} \right), & v\geq v_*\,.
\end{cases}
\eeq

\subsection{MSW models}

Our next examples are the 2d generalizations of the SYK model considered by Murugan, Stanford and Witten in \cite{Murugan:2017eto}. These Lorentz invariant models flow to  conformal field theories in the IR at large $N$ and after disorder averaging. It is an interesting open problem to determine their fate for one realization of the disorder and at finite $N$. There are two similar models considered in \cite{Murugan:2017eto}. There is a bosonic model with an $N$ component boson $\phi^i$ and action
\beq
I=\int d^2 x \left[ \frac{1}{2} (\partial \phi^i)^2 + \sum_{i_1,...,i_q} J_{i_1...i_q}\phi^{i_1}\cdots \phi^{i_q} \right],
\eeq
with $J_{i_1...i_q}$ independent Gaussian random variables with zero mean and variance $\langle J_{i_1...i_q}^2 \rangle=\frac{J^2}{q N^{q-1}}$. This model is unfortunately not stable; the scalar potential has negative directions. Nevertheless, at large $N$ it still has a conformal solution in the IR, but the corresponding CFT has some complex scaling dimensions. There is a version of the model with a similar action, where $\phi^i$ are replaced by superfields, resulting in a model with $\mathcal{N}=1$ supersymmetry. This model is stable and flows to a genuine 2d CFT in the IR (up to the caveat about disorder averaging and large $N$).

For both of these models, the OTOC can be calculated, and the growing contribution has the form \cite{Murugan:2017eto}
\beq
\label{eq:MSWRegge}
f(t,x) \approx 1 -\frac{1}{N} \int \frac{dp}{2\pi}\, g(p)\, e^{[j(p)-1]t+i px},
\eeq
where $g(p)$ and $j(p)$ are known functions and we set $\beta=2\pi$. In fact, \eqref{eq:MSWRegge} is also the Regge limit of the flat space\footnote{In two dimensions, the thermal OTOC is related to the flat space correlator via the exponential map, see \cite{Roberts:2014ifa}.} four point function of the IR CFT, and $j(p)$ is its leading Regge trajectory. This example is a special case of the general discussion of the VDLE on hyperbolic space presented in sec.~\ref{sec:Regge}.

In both models, the function $j(p)$ is given by an implicit equation of the form
\beq
\label{eq:MSWtrajectory}
\kappa(j(p),p)=1,
\eeq
where $\kappa(j,p)$ is the eigenvalue of the four point ladder kernel. One may write it explicitly in the two models of \cite{Murugan:2017eto} as
\bea
\label{eq:MSWeigenval}
\kappa(j,p)&=
\begin{cases}
k_{\rm bosonic}\left[\frac{j+1}{2}-i\frac{p}{2},-\frac{j-1}{2}-i\frac{p}{2}\right]\\
k_{\rm SUSY}\left[\frac{j}{2}-i\frac{p}{2},-\frac{j-1}{2}-i\frac{p}{2}\right]
\end{cases}, \\
k_{\rm bosonic}[h,\bar h]&=\frac{\Delta(2-\Delta)\Gamma(2-\Delta)^2\Gamma(\Delta+h-1)\Gamma(\Delta-\bar h)}{\Gamma(1+\Delta)^2\Gamma(1+h-\Delta)\Gamma(2-\bar h -\Delta)},\\
k_{\rm SUSY}[h,\bar h]&=-\frac{\Gamma(1-\Delta)^2}{\Gamma(\Delta-1)\Gamma(\Delta+1)} \frac{\Gamma(h+\Delta-\frac{1}{2})}{\Gamma(h-\Delta+\frac{1}{2})} \frac{\Gamma(-\bar h +\Delta)}{\Gamma(1-\bar h-\Delta)},
\eea
where $\Delta=1/q$.

\begin{figure}[!h]
\begin{center}
\includegraphics[width=0.45\textwidth]{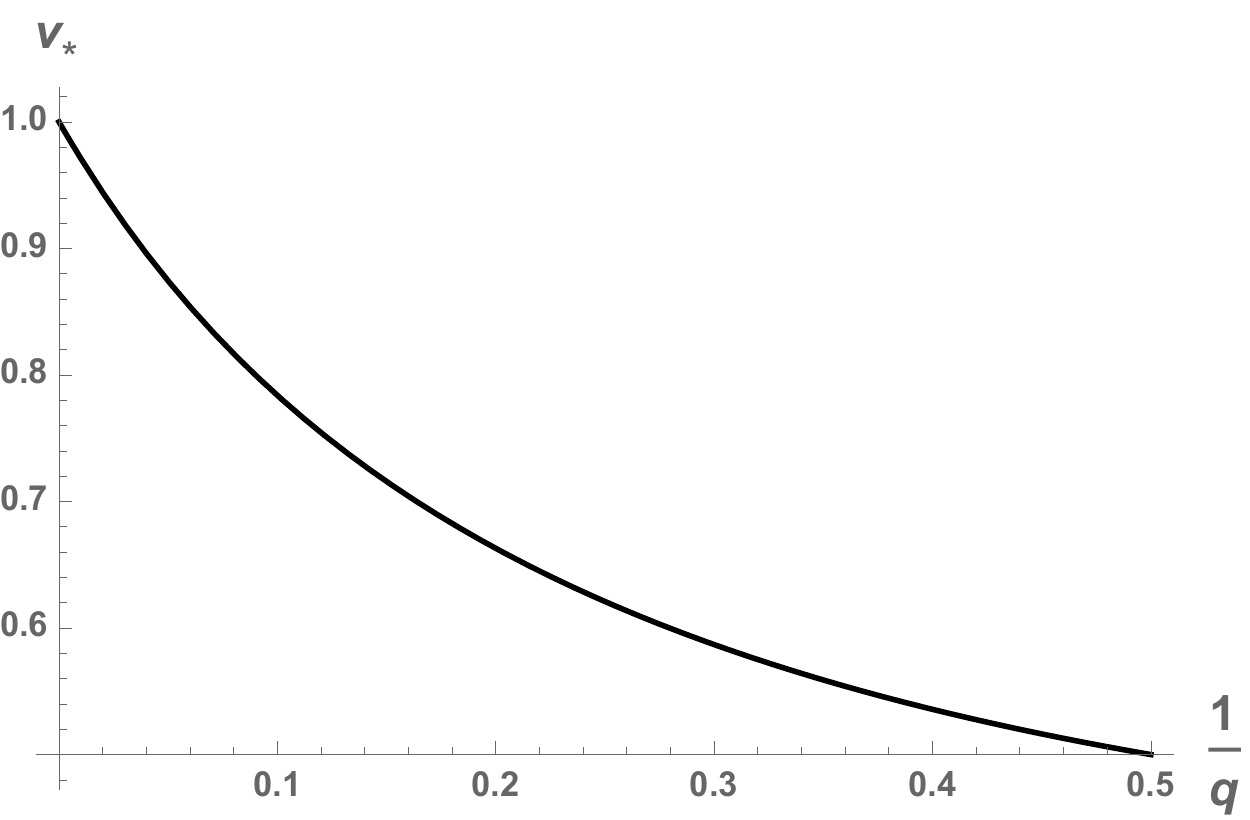} $\quad$ \includegraphics[width=0.45\textwidth]{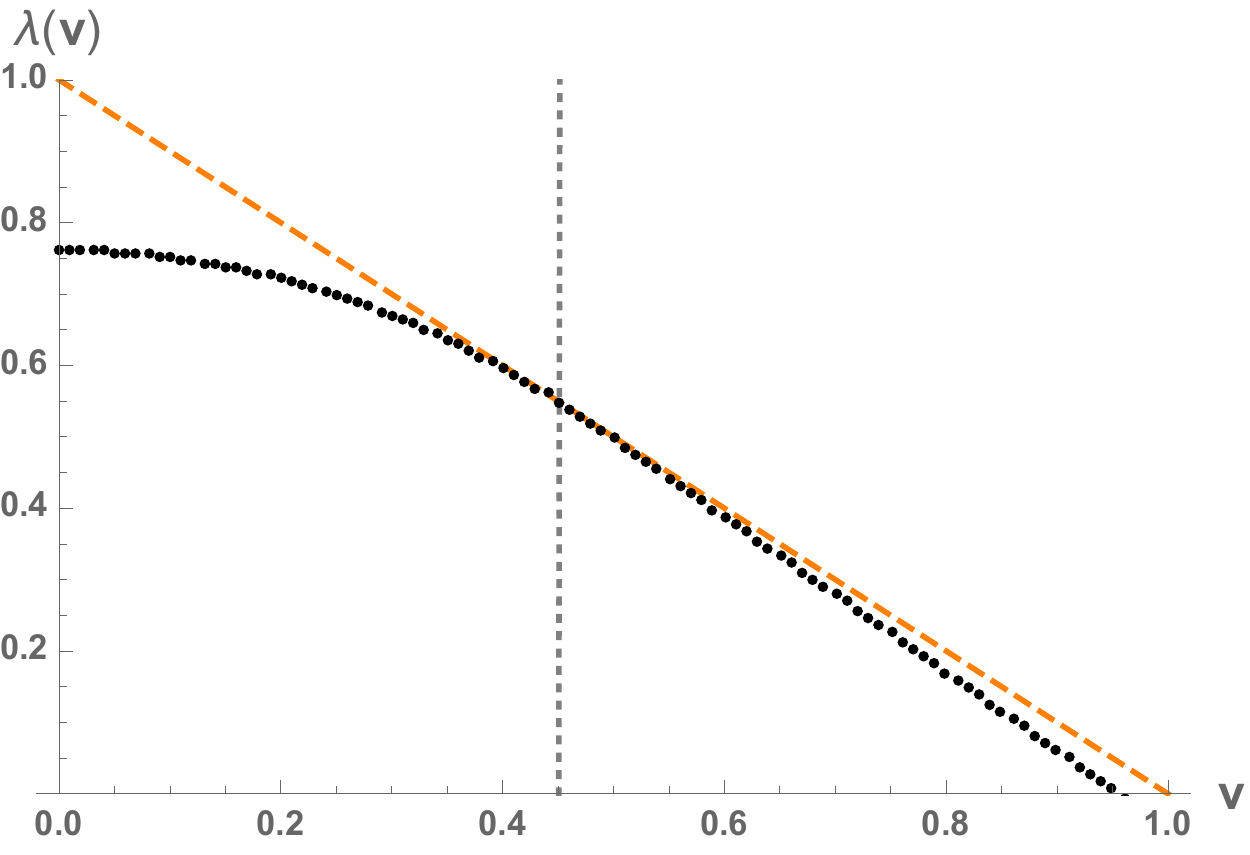}
\caption{Left: critical velocity $v_*$ in the bosonic MSW model. Right: The black dots are the saddle point contribution to the VDLE, the orange, dashed line is the contribution of the pole at $p=i$, for the bosonic MSW model and $q^{-1}=0.75$. (In order to see definite deviation between the black dots and the orange line, we use an unphysical value of $q$ in this figure.) They exchange dominance at the critical velocity, drawn gray, dotted. Virtually identical plots can be produced for the SUSY version of the model.  \label{fig:MSW}}
\end{center}
\end{figure}

To obtain the VDLE, we set $x=vt$ in \eqref{eq:MSWRegge} and evaluate the integral by saddle point at large $t$. As before, there is a pole coming from $g(p)$ when $j(p)=2$, or $p=\pm i$ \cite{Murugan:2017eto}. As we increase $v$, we eventually cross the pole at $p=i$, and after this, the integral is dominated by the pole, see Fig.~\ref{fig:exchangeofdom}. The critical velocity can be explicitly determined by equating the location of the saddle point with the pole and using \eqref{eq:MSWtrajectory}
\beq
v_*=ij'(i)=-i\frac{\partial_p \kappa}{\partial_j \kappa}\Big\vert_{j=2,p=i}.
\eeq
In the case of the supersymmetric model, it has a remarkably simple analytic form
\beq
v^{\rm SUSY}_* = \frac{2\Delta-1}{1-2\Delta+2\Delta(\Delta-1)\pi \cot \pi \Delta}.
\eeq
One interesting feature of $v_*$ is that it approaches the speed of light $v=1$ when $q\rightarrow \infty$, see left of Fig.~\ref{fig:MSW}. This is a weakly coupled limit of these models and we see that the region near the light cone with maximal chaos persists for all values of $q$. We cannot write a closed formula for the saddle point contribution to the VDLE, but we can evaluate it numerically, this is shown on the right of Fig.~\ref{fig:MSW}.

\subsection{$\mathcal{N}=(0,2)$ SYK}

\begin{figure}[!h]
\begin{center}
\includegraphics[width=0.3\textwidth]{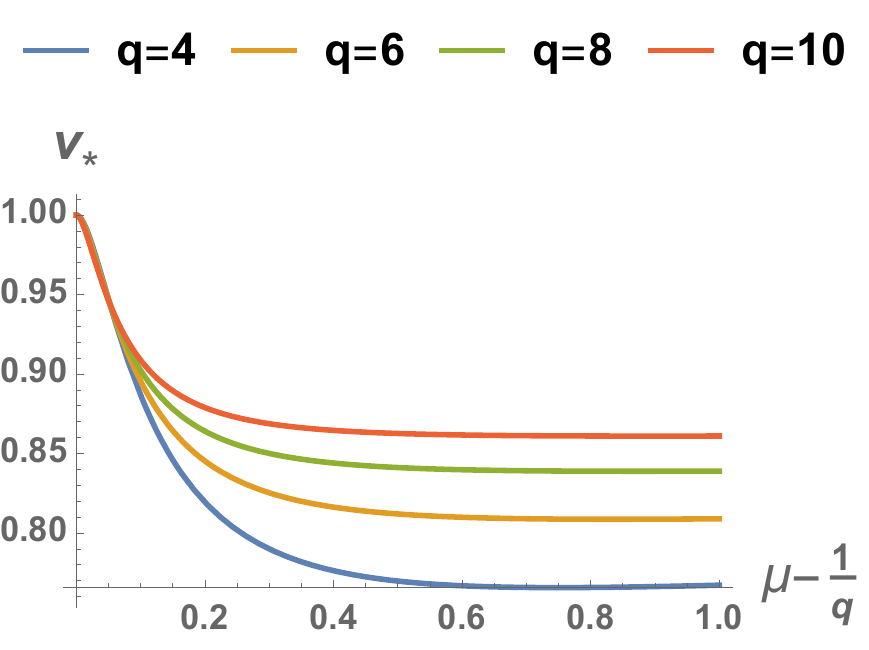}  \includegraphics[width=0.3\textwidth]{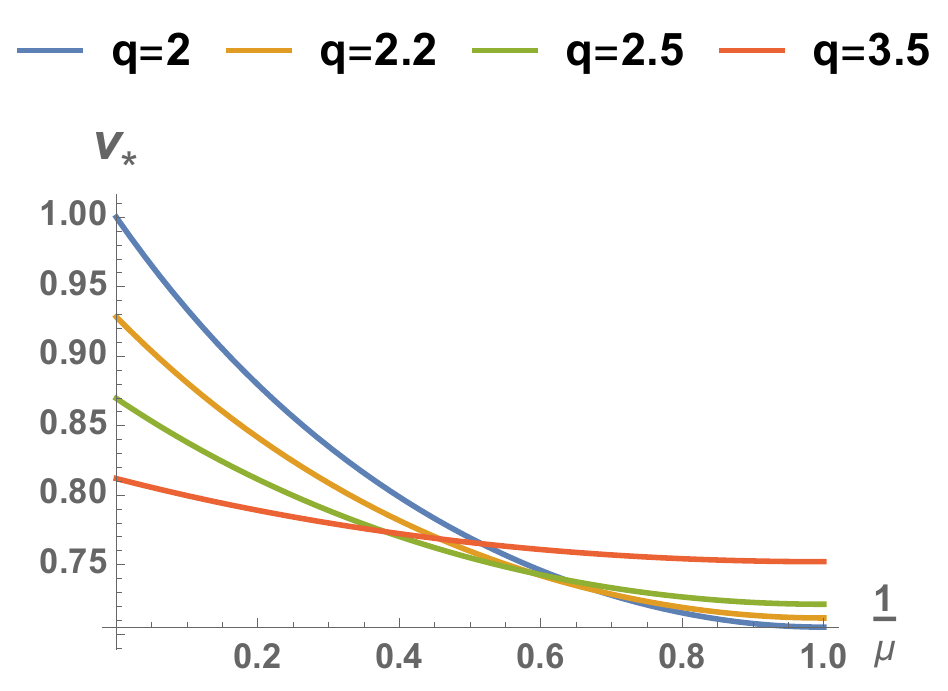}  \includegraphics[width=0.3\textwidth]{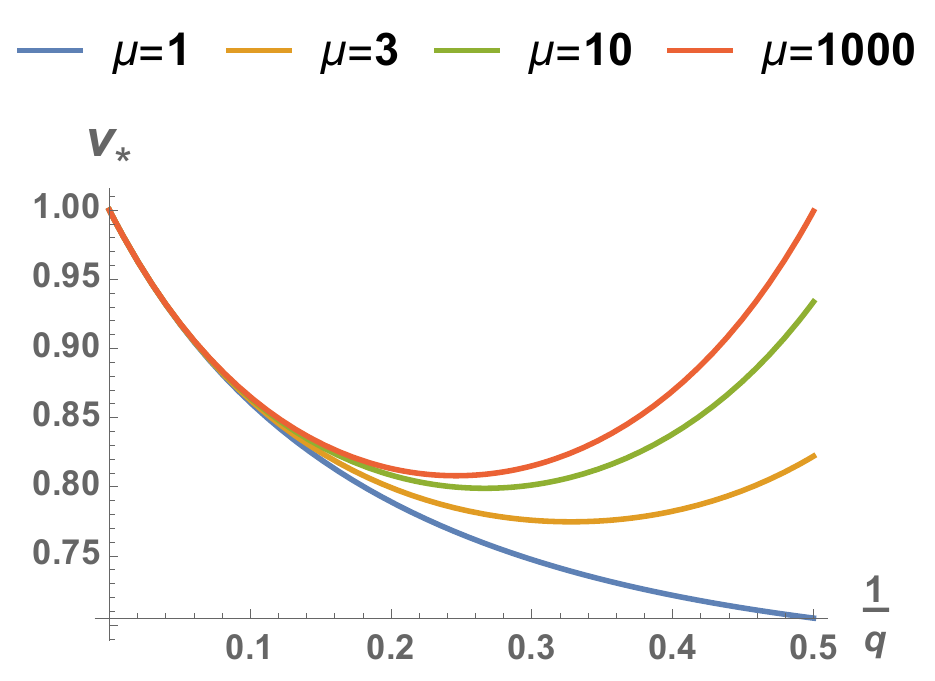}
\caption{Left and middle: $\mu$ dependence of the critical velocity in $\mathcal{N}=(0,2)$ model with various $q$. Right: $q$ dependence of the critical velocity at different values of $\mu$. \label{fig:Peng}}
\end{center}
\end{figure}

There is a variant of the MSW model with $\mathcal{N}=(0,2)$ chiral supersymmetry due to Peng \cite{Peng:2018zap}. These models flow to a two-parameter family of IR CFTs characterized by $q$ that is present in all SYK models and a new parameter $\mu$ that takes values between $q^{-1}$ and $\infty$. At these two extremal values, the model develops higher spin symmetry. The weakly coupled limits are $(\mu\rightarrow q^{-1},q\text{ arbitrary})$, $(\mu\rightarrow \infty,q=2)$, and $(\mu\text{ arbitrary },q\rightarrow \infty)$. The Lyapunov exponent vanishes at these points. In this model, there is still an implicit equation like \eqref{eq:MSWtrajectory} defining the leading Regge trajectory, but it is much more complicated than \eqref{eq:MSWeigenval}.\footnote{We omit the formulas here, they can be found in \cite{Peng:2018zap}, e.q. (3.42)-(3.56).} Nevertheless, we can plot the critical velocity of the model as a function of the parameters $\mu$ and $q$, see Fig.~\ref{fig:Peng} for some examples. We have $v_*\rightarrow 1$ continuously in all the weakly coupled limits.

\subsection{Chiral SYK}
\label{sec:chiralSYK}

\begin{figure}[!h]
\begin{center}
\includegraphics[width=0.3\textwidth]{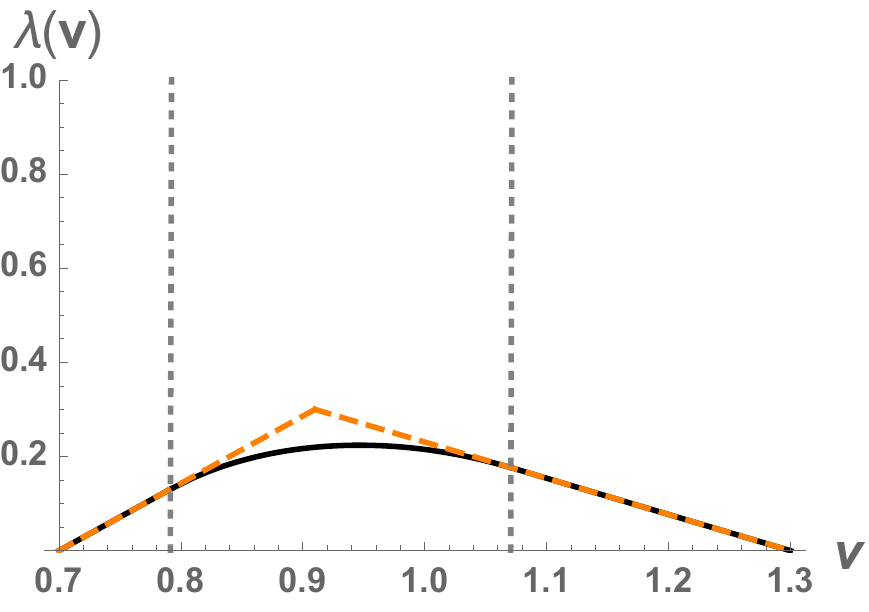}\includegraphics[width=0.3\textwidth]{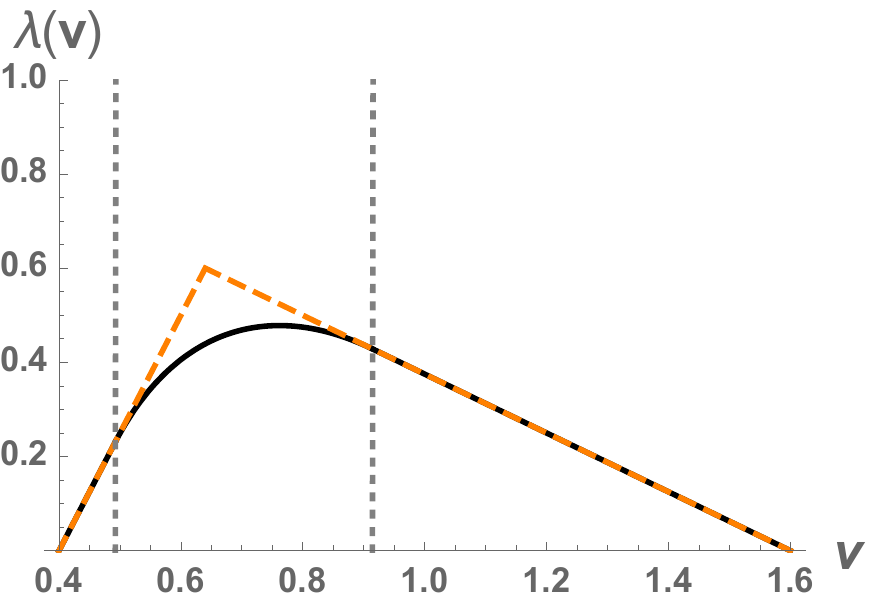}\includegraphics[width=0.3\textwidth]{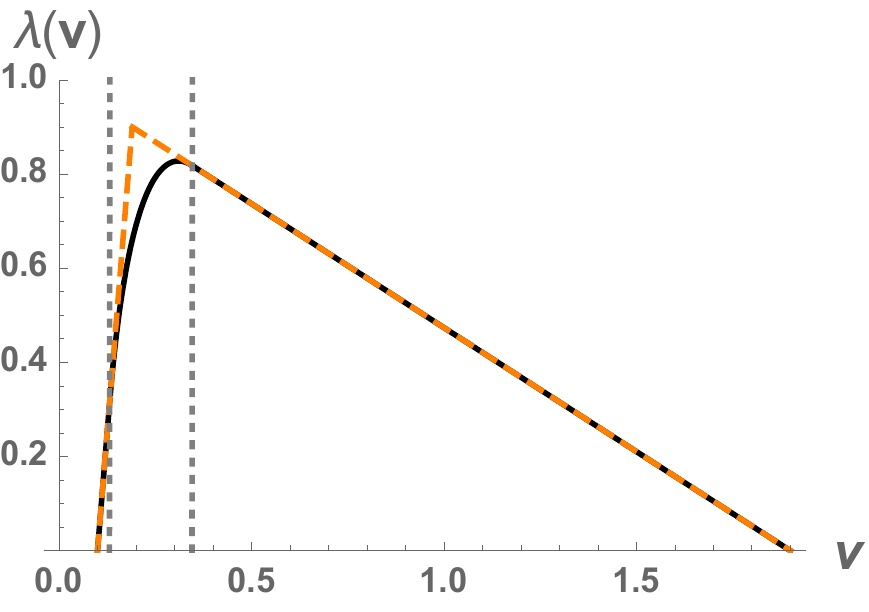}
\caption{Velocity dependent Lyapunov exponent (black solid) against the bound \eqref{BoundSummary} (orange dashed) along with the critical velocities (gray dotted) in the chiral SYK model (for $\beta=2\pi$). The couplings from left to right are $\mathcal{J}=0.3,0.6,0.9$.  \label{fig:chiralSYK}}
\end{center}
\end{figure}

Another 2d generalization of the SYK model is constructed by Lian, Sondhi and Yang in \cite{Lian:2019axs}. This model is non-relativistic, and a very interesting feature is that it has an asymmetric butterfly cone, with upper and lower butterfly velocities $v_B^+$ and $v_B^-$. Similarly to the previous examples, the growing contribution to the OTOC is given by an integral,
\beq
\label{eq:chiralSYKotoc}
f(t,x) \supset \int dp \frac{e^{\frac{2\pi}{\beta}[\varkappa(p)t+i px]}}{\cos \frac{\pi \varkappa(p)}{2}},
\eeq
where $\varkappa(p)$ is a function determined by solving an equation involving the eigenvalues of the retarded ladder kernel, similarly to the previous examples. It reads as
\beq
\varkappa(p)=\frac{\mathcal{J} \sqrt{3 \left(1-\mathcal{J}^2\right)+\left(\mathcal{J}+i \left(1-\mathcal{J}^2\right) p\right)^2}-i \left(1-\mathcal{J}^2\right) p-\mathcal{J}}{1-\mathcal{J}^2},
\eeq
where $\mathcal{J}$ is a marginal coupling with values $0\leq \mathcal{J} \leq 1$. Again, there is an exchange of dominance between the saddle point contribution and contribution of poles. The actual situation is a bit more elaborate than the one described in Fig.~\ref{fig:exchangeofdom} this time, because $\varkappa(p)$ has branch cuts and the steepest descent contour runs between the two branches. Nevertheless, the VDLE can be evaluated \cite{Lian:2019axs} and the upshot is that there is both a lower and an upper critical velocity $v_*^\pm$ such that for $v<v_*^-$ and $v>v_*^+$, the VDLE is ballistic. The critical and butterfly velocities are
\beq
v_*^\pm = \frac{2-2\mathcal{J}^2}{2\mp \mathcal{J}}, \quad \quad v_B^\pm = 1\pm \mathcal{J}.
\eeq
The model obeys the bound \eqref{BoundSummary} valid for asymmetric butterfly cones, and saturates it for $v<v_*^-$ and $v>v_*^+$. We plot the VDLE of the model for some values of the couplings on Fig.~\ref{fig:chiralSYK}.

\subsection{Ladder identity of Gu and Kitaev}

In all the SYK-like examples, the saturation of the bounds \eqref{eq:VDLObound1}-\eqref{eq:VDLObound2} beyond the critical velocity was a consequence of an exchange of dominance between a saddle point and a pole in an integral computing the OTOC. The presence of this pole is generic in theories where the four point function is a sum of conformal ladders, as a consequence of the ladder identity derived by Gu and Kitaev in \cite{Gu:2018jsv}. For our purposes, this identity essentially states that the contribution of a growing mode of exponent $\varkappa$ to the OTOC is
\beq
\sim \frac{1}{N\cos \frac{\pi \varkappa}{2}} e^{\varkappa t},
\eeq
and therefore if $\varkappa$ is an analytic function of some mode parameter that we are integrating over, there is always a pole at $\varkappa=1$. We will see in the next sections, that a very similar mechanism is at play both for generic large $N$ CFTs on hyperbolic space (which include thermal 2d CFTs on the line), and strongly coupled holographic CFTs in flat space. The presence of the pole is intimately related to the bound \eqref{eq:VDLObound1} on the Legendre transform of the VDLE. Indeed, in an integral over modes of the generic form \eqref{eq:chiralSYKotoc}, the saddle point contribution to $\lambda(v)$ is the Legendre transform of $\varkappa(p)$. So we can think of the bound \eqref{eq:VDLObound1} as saying that the saddle point can only dominate as long as $|\varkappa(p)|\leq 1$. When $|\varkappa(p)|$ reaches one, a pole takes over. On Fig.~\ref{fig:boundvssaddle} we explain how to understand this in terms of the direct bound \eqref{eq:VDLObound2} on the VDLE. We will also return to this question in the context of the Regge limit in sec.~\ref{sec:Regge}.

\subsection{Stringy corrections to the gravity result}

Using pure Einstein gravity in the bulk, the AdS/CFT correspondence predicts the OTOC \cite{Shenker:2013pqa,Roberts:2014isa,Shenker:2014cwa}
\es{EinsteinRes}{
f(t,{\bf x}) \approx 1-\# G_N e^{\frac{2\pi}{\beta}\left( t-\frac{ |{\bf x}|}{v_B} \right)}, \quad v_B=\sqrt{\frac{d}{2(d-1)}},
}
which leads to the velocity dependent Lyapunov exponent
\es{LambdaGrav}{
\lambda({\bf v})=\frac{2\pi}{\beta}\left( 1-\frac{ |{\bf v}|}{v_B} \right),
}
saturating the bound \eqref{eq:VDLObound2}. This corresponds to the case of infinite coupling in the CFT. In AdS/CFT finite coupling corrections come both from higher derivative corrections in the supergravity action and exchange of strings, since the dimensionless coupling of the CFT corresponds to some power of $\ell_{\rm AdS}/\ell_{\rm string}$ in the bulk. The former only changes the value of $v_B$, but does not change the function \eqref{LambdaGrav} \cite{Roberts:2014isa,Maldacena:2015waa,Mezei:2016wfz,Grozdanov:2018kkt}.  Stringy corrections to the OTOC were calculated by Shenker and Stanford in \cite{Shenker:2014cwa}. They have found the growing contribution
\beq
f(t,{\bf x}) \supset \# G_N \int \frac{d^{d-1}{\bf k}}{(2\pi)^{d-1}}G({\bf k}^2+\mu^2)e^{i {\bf k}\cdot {\bf x}+\frac{2\pi}{\beta}t\left[1-\frac{\ell_{\rm string}^2}{2r_0^2}({\bf k}^2+\mu^2)\right]},
\eeq
where $\mu=2\pi \beta^{-1}v_B^{-1}$,\footnote{The $v_B$ in this relation is expected to be the one after higher derivative corrections have been taken into account, not just the Einstein gravity result  \eqref{EinsteinRes}.} $r_0=4\pi\ell_{\rm AdS} d^{-1}\beta^{-1}$ is the horizon radius, and $G(\xi)$ is a known function, whose only important feature in the above integral is that is has a simple pole at $\xi=0$. As discussed already in \cite{Shenker:2014cwa} for small $|{\bf x}|$ and large $t$, the integral is dominated by a saddle point, while for large $|{\bf x}|$ it is dominated by the pole. This is made more transparent with the use of the VDLE: if we substitute ${\bf x}={\bf v}t$, we can always evaluate the integral by saddle point, but the steepest descent contour crosses the pole at the critical velocity, and after this, the pole dominates, see Fig.~\ref{fig:exchangeofdom}. This is the same mechanism that we have seen for SYK models. The critical velocity is when the saddle and the pole coincide
\beq
v_*=\frac{d^2}{4 v_B} \left( \frac{\ell_{\rm string}}{\ell_{\rm AdS}}\right)^2,
\eeq
that is, proportional to some inverse power of the coupling in the boundary. The VDLE is
\beq
\lambda(v)=
\begin{cases}
 \frac{\pi}{\beta} \left( 2-\frac{v_*}{v_B}-\frac{v^2}{v_* v_B}\right) , & v<v_*\\
 \frac{2\pi}{\beta}\left(1-\frac{v}{v_B} \right), & v\geq v_*
\end{cases}.
\eeq
Remarkably, when expressed with the velocities $v_B$ and $v_*$, this is exactly the same form as in the SYK chain, \eqref{eq:VDLEchain}.

\subsection{When the pole does not dominate}

We note that while the examples we have discussed in this section do not realize it, there exists a scenario, where within the butterfly cone the pole contribution never dominates. In this case, we have $v_*>v_B$. This can happen at weak coupling, for instance, we will argue in sec.~\ref{sec:Regge} that it happens for large $N$ CFTs on Rindler space in $d>2$ that are sufficiently weakly coupled. A particular example of this is $\mathcal{N}=4$ SYM at weak coupling. On the other hand, we will see that it cannot happen for a 2d CFT. Another example of this scenario is the SYK chain with bilinear fermion couplings between sites, or the so called $t-U$ model \cite{Song:2017pfw}, which has a Fermi liquid phase where it displays $v_*>v_B$, even in $d=2$ \cite{Guo:2019csw}. To illustrate this behavior in a simple example, we can formally extrapolate the SYK chain result \eqref{SYKChainInt} outside the regime of its validity to such values of parameters for which $1<{12\pi \al\ov \beta J}<2$. Then inside the butterfly cone we have only the first line of  \eqref{eq:VDLEchain} expressed as :
\es{eq:VDLEchain2}{
\lambda(v)=
 \frac{2\pi}{\beta} \left( 1-{6\pi \al\ov \beta J}-\frac{v^2}{48\pi^2 \al D/ \beta^2 J}\right)\,.
}
Note that this $\lam(v)$ does not saturate the bound \eqref{eq:VDLObound2} inside the butterfly cone (only at $v_B<v_*$).\footnote{To be completely explicit, $v_*={12\pi \al\ov \beta J}\, \sqrt{2\pi D\ov \beta}$ just as in \eqref{vsvbSYK}, but now $v_B=\sqrt{{\beta J\ov 6\pi \al}-1}\,v_*$, which in the parameter regime of interest gives $v_B<v_*$. }

%\pagebreak
\section{Conformal Regge theory}
\label{sec:Regge}

\subsection{Rindler OTOC}

\begin{figure}[!h]
\begin{center}
\includegraphics[width=0.5\textwidth]{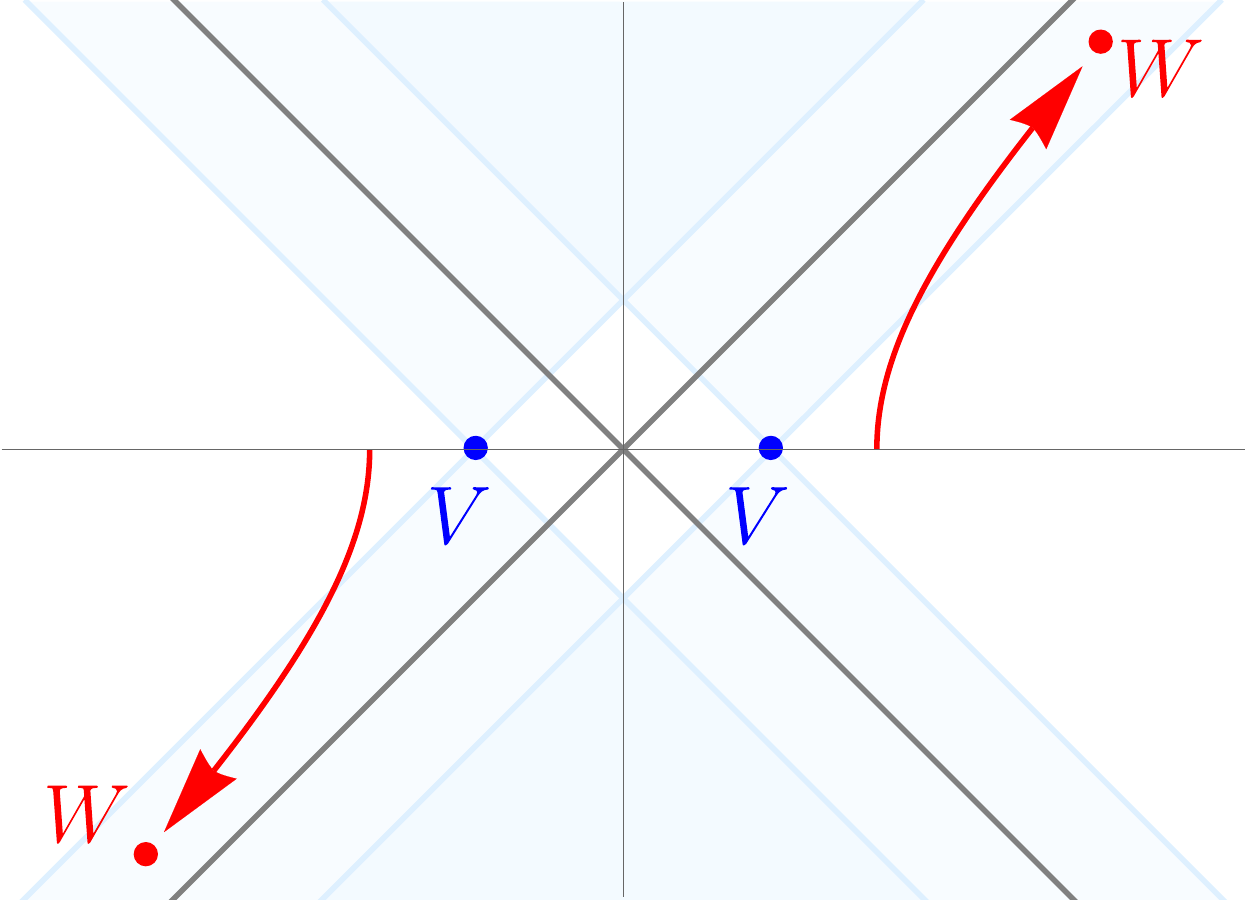}
\caption{Operator insertions in the flat space correlator. Two operators are in the left and two operators are in the right Rindler wedge. The $W$ operators follow the red lines as we increase Rindler time. When they enter the blue region, the cross ratio $\bar z$ moves to the second sheet.\label{fig:operators}}
\end{center}
\end{figure}

Let us consider the \textit{vacuum} four point function in a $d$ dimensional conformal field theory
\beq
\label{eq:flatspace4pt}
f(z,\bar z)=\frac{\langle V(x_1)W(x_2)W(x_3)V(x_4)\rangle }{\langle V(x_1)V(x_4)\rangle \langle W(x_2)W(x_3) \rangle},
\eeq
depending only on the conformal cross ratios defined through
\bea
z \bar z &= u, && &(1-z)(1-\bar z)=v, \\
u&=\frac{x_{12}^2x_{34}^2}{x_{13}^2x_{24}^2}, && &v=\frac{x_{14}^2x_{23}^2}{x_{13}^2x_{24}^2}.
\eea
We will be interested in the (Lorentzian) alignment (Fig.~\ref{fig:operators})
\bea
x_1=(0,-1,0,...), && x_2=(-t,-y,0,...), && x_3=(t,y,0,...), && x_4=(0,1,0,...),
\eea
for which we have
\bea
z&=\left( \frac{y-1-t}{y+1-t}\right)^2, && \bar z&=\left( \frac{y-1+t}{y+1+t}\right)^2.
\eea

What does this setup have to do with thermal OTO correlators such as \eqref{eq:OTOCdef}? The answer is that we can interpret this correlator as such a thermal OTOC when $t<y$. In this case, two operators are in the left Rindler wedge and two operators are in the right Rindler wedge. Picking Rindler coordinates
\bea
t=U \sinh T, && y=U \cosh T,
\eea
we find a thermal space with Euclidean periodicity $T\sim T+2\pi i$. In fact, the metric becomes
\beq
ds^2=U^2\left( -dT^2+\frac{dU^2+dx_{\perp}^2}{U^2} \right),
\eeq
which is conformal to $\mathbb{R}\times \mathbb{H}^{d-1}$. Therefore, we can interpret \eqref{eq:flatspace4pt} as a thermal correlator on the hyperboloid $\mathbb{H}^{d-1}$ with inverse temperature $\beta=2\pi$. The operator insertions become:
\beq
\langle V(i\pi,1)W(T+i\pi,U)V(0,1)W(T,U)\rangle_{S^1\times \mathbb{H}^{d-1}},
\eeq
which, in case we send $T\rightarrow T-i\epsilon$ to resolve the ordering of the $V$ and $W$ operators, is the value of the OTO correlator, defined in \eqref{eq:OTOCdef}, on the boundary of the strip (since we would have $T=t+i(\beta/4-\epsilon)$ in the variables of \eqref{eq:OTOCdef}). This $\epsilon$ prescription translates to standard time ordering $t\rightarrow t(1-i\epsilon)$ in the flat space picture \eqref{eq:flatspace4pt}. This is because Rindler time flows in opposite directions in the two wedges.

The cross ratios can be written in terms of Rindler coordinates as
\bea
z&=\left( \frac{1-e^{-T} U}{1+e^{-T} U}\right)^2, && \bar z&=  \left( \frac{1-e^{-T} U^{-1}}{1+e^{-T} U^{-1}}\right)^2.
\eea
We are interested in the Lyapunov regime, which is $T\gg 2\pi$, in which case
\bea
\label{eq:cross1}
z=1-4U e^{-T} + \cdots,\\
\bar z= 1-\frac{4}{U} e^{-T} + \cdots,
\eea
So both $z$ and $\bar z$ approach 1. Note that starting from the Euclidean sheet, this approach happens after continuing to the second sheet $ z\rightarrow e^{2\pi i} z$. This is because in the Lorentzian picture, the $W$ operators cross the light cones of the $V$ operators as we increase Lorentzian Rindler time $T$ from zero to a large value (see Fig.~\ref{fig:operators}, we assume $U>1$ without loss of generality). Note that
\beq
\frac{1-z}{1-\bar z} \approx U^2,
\eeq
so if we fix $U$ while sending $\bar z\rightarrow 1$, this is called the Regge limit, studied in great detail in \cite{Cornalba:2007fs,Costa:2012cb,Kravchuk:2018htv}. This is the case when we do not scale the separation of operators with $T$ and the formula \eqref{eq:OTOCdecay} for the OTOC is appropriate. 

We would like to instead scale $U$ with $T$. It is not entirely obvious how to do this when the spatial manifold is not a linear space, but assuming that the CFT couples covariantly to the background metric, we must replace $|{\bf x}|$ with the geodesic distance $\rho$ between $V$ and $W$ on the spatial manifold. (This is also what we do in $d=2$, where we are just working with a thermal CFT on a line.) The formula \eqref{eq:OTOCVDLE} defining the VDLE then becomes
\es{HypOTOC}{
f=1-\epsilon\, e^{\lambda\le(\frac{\rho}{t}\ri)\,t} + \cdots.
}
In the hyperbolic metric, the spatial distance $\rho$ between $V$ and $W$ is written as 
\beq
\label{eq:rhodef}
\rho = \int^U_1 \frac{dU'}{U'}=\log U.
\eeq
We note that in the recent holographic computation of \cite{Ahn:2019rnq} the OTOC indeed obeyed the scaling \eqref{HypOTOC}, but we will justify this ansatz for any large $N$ CFT in the next section.
 We then want to send $T\rightarrow \infty$ and scale $\rho=v T$. 
This is a limit where $\bar z \rightarrow 1$ with
\beq
\frac{1-z}{(1-\bar z)^\frac{1-v}{1+v}}=\text{fixed}.
\eeq
For $v=0$ this is the Regge limit as before. For $v=1$, we need to fix $z$ while, $\bar z \rightarrow 1$, which is called the light cone limit. These two limits are dominated by different physics, as we will soon review, and the above limit interpolates between them.

\subsection{VDLE and the leading Regge trajectory}

\subsubsection*{Resummation in the Regge limit}

The $z,\bar z \rightarrow 1$ limit naively looks like an OPE limit. However, since the cross ratio $z$ encircles a branch cut at $z=0$, this OPE expansion is no longer convergent. This can be seen by using the known monodromy of global conformal blocks, leading to the behavior $\sim [(1-z)(1- \bar z)]^{\frac{1-J}{2}}$ of each term of spin $J$ in the OPE in the $z,\bar z \rightarrow 1$  limit \cite{Cornalba:2007fs}.

This is a standard feature of the Regge limit, and there is a standard way of dealing with it, called the Sommerfeld-Watson resummation, worked out in detail in \cite{Cornalba:2007fs,Costa:2012cb,Kravchuk:2018htv}.  The idea is to write $f(z,\bar z)$ in \eqref{eq:flatspace4pt} as a conformal partial wave decomposition, consisting of an integral over principal series $\Delta=d/2+i\nu$ and a sum over spin. In such a partial wave decomposition, the OPE coefficients are replaced by the conformal partial amplitudes $b_J(\nu^2)$, while the conformal blocks by conformal partial waves (these are just combination of the block and a shadow block). The conformal partial amplitudes are required to be meromorphic functions of $\nu$ for $\text{Im}\nu<0$ with isolated poles. The poles are such that the regular OPE is reproduced when the contour is pulled down from the real $\nu$ axis. For this, $b_J(\nu^2)$ needs to have poles when $\Delta=d/2+i\nu$ hits physical operators in the spectrum, but it also needs to have poles arranged in a way that cancels any contribution from unwanted ``kinematical" poles coming from the partial wave. In addition, $b_J(\nu^2)$ admits a nice analytic continuation in $J$, as shown by Caron-Huot \cite{Caron-Huot:2017vep}. This continuation is nice because it is bounded for large $\text{Re}\,J$ \cite{Kravchuk:2018htv}. This allows one to perform the Sommerfeld-Watson resummation, which consists of writing the $J>0$ part of the $J$ sum in the conformal partial wave decomposition as a contour integral over $J$, with the contour encircling positive integers, and then pull this contour to $\text{Re}\,J<1$, where the partial waves stay bounded in the Regge limit. Doing so, one picks up contributions from other non-analyticities in the right $J$ half plane, which determine the form of the growth in the Regge limit. It is generically assumed that the rightmost non-analyticity is a pole, located at some $J=j(\nu)$. The function $j(\nu)$ is called the leading Regge trajectory. Then, the growing part of the four point function is given by an integral of the residue over this Regge trajectory \cite{Cornalba:2007fs,Costa:2012cb,Kravchuk:2018htv}\footnote{Our $\rho$ is the same as the one used by \cite{Costa:2012cb}, and their $\sigma$ is our $4e^{-T}$.}
\beq
\label{eq:SWresummed}
f(z,\bar z) \supset \int d\nu g(\nu)e^{[j(\nu)-1]T}\Omega_{i\nu}(\rho),
\eeq
where the definition of $T$ and $\rho$ in terms of the cross ratios is in \eqref{eq:cross1}, \eqref{eq:rhodef}, and $\Omega_{i\nu}(\rho)$ is a harmonic function on $\mathbb{H}^{d-1}$ given in Appendix C of \cite{Costa:2012cb}. We will only need its form when $\rho \rightarrow \infty$, in which case
\beq
\label{eq:Omega}
\Omega_{i\nu}(\rho)= \text{Re}\Big[ \frac{\nu}{2} \pi^{-1-\frac{d}{2}} \sinh \pi \nu \Gamma(-i \nu)\Gamma(i\nu+\frac{d}{2}-1) e^{\rho(1-\frac{d}{2}-i\nu)}\Big]\Big(1+O(e^{-\rho}) \Big).
\eeq
The function $g(\nu)$ contains contribution from the partial amplitude $b_J(\nu^2)$ (more precisely, its residue on the Regge trajectory $J=j(\nu)$), containing the dynamical data, but also other kinematical factors that descend from the partial waves. The precise form of it will be irrelevant for us, but it will be important that it contains an explicit factor of $\left(\sin \frac{\pi j(\nu)}{2} \right)^{-1}$ coming from the kinematical factors. Its explicit form is written down in \cite{Costa:2012cb,Costa:2017twz}. 

Some useful properties of the Regge trajectory $j(\nu)$ are the following. First, we have to distinguish between two notions of Regge trajectories: the large $N$ one only containing single-trace operators, and the ``real'' Regge trajectory; for our applications we will be interested in the former.  We will assume that $j$ is analytic in $\nu$ in some sufficiently big region containing $\nu=0$. For imaginary $\nu=-ir$, we have $\Delta=d/2+r$. For various positive values of $r$, we hit a physical dimension in the spectrum, in this case the value of $j(\nu)$ must be the highest spin corresponding to that dimension (this is bounded by unitarity).\footnote{This is because $j(\nu)$ was defined to be the rightmost pole of a function that contains the conformal partial amplitude, which must have poles at physical operators.} Moreover, there is a symmetry under $r\rightarrow -r$ corresponding to the ``shadow" symmetry $\Delta \rightarrow d-\Delta$. This makes $j$ an even function, ensuring that it is real along both the real and imaginary axis, and that $\nu=0$ is a saddle point of $j$. In addition, it can be shown for the ``real" Regge trajectory that along imaginary $\nu$, $j$ is convex, while along real $\nu$ it is concave \cite{Komargodski:2012ek,Costa:2017twz}. This result has not been proven for the large $N$ Regge trajectory that is relevant for chaos, we will nevertheless assume that it also holds for the large $N$ Regge trajectory.

\subsubsection*{Lyapunov exponent}

Now let us move on to discuss the Lyapunov exponent. For large $T$ and small $\rho$, we evaluate the integral \eqref{eq:SWresummed} by saddle point. Since we have seen that $j$ has a saddle at $\nu=0$, this is giving an exponential behavior $\sim e^{[j(0)-1]T}$. $j(0)$ is called the Regge intercept and we can identify it with the Lyapunov exponent, $\lambda_L=j(0)-1$. The MSS chaos bound in this case states that $j(0)\leq 2$.\footnote{This is a bound on the large $N$ Regge intercept. The ``true" Regge intercept must satisfy $j(0)\leq 1$ because the four point function must be bounded in the Regge limit \cite{Caron-Huot:2017vep}.}

We are now in position to discuss the velocity dependent Lyapunov exponent, $\lambda(v)$, obtained by putting $\rho=vT$ in $f(z,\bar z)$. For large $T$, the effect of this is to shift the saddle point in $\nu$. Namely, putting \eqref{eq:Omega} in \eqref{eq:SWresummed} and using the symmetry of $g(\nu)$ and $j(\nu)$ under the shadow transform $\nu \rightarrow -\nu$  we obtain
\beq
\label{eq:Reggesaddle}
f \sim \int d\nu g(\nu)e^{[j(\nu)-1-(i\nu+d/2-1) v]T}\,,
\eeq
so we are adding a linear function to $j(\nu)$, whose effect is to shift the saddle from $\nu=0$ in the imaginary $\nu$ direction.\footnote{A very similar discussion can be found in \cite{Costa:2017twz}, who uses the ability to shift the saddle to put bounds on analytically continued OPE data on the leading Regge trajectory from AdS unitarity. Here, we merely focus on the exponent and do not discuss its coefficient. Another key difference is that the unitarity bounds considered in \cite{Costa:2017twz} require restricting to the real part of $g(\nu)$, from which the poles at even $j$ coming from $\left(\sin \frac{\pi j(\nu)}{2} \right)^{-1}$ turn out to cancel. For us, the pole at $j=2$ will be crucial for the bound \eqref{eq:VDLObound1} to be obeyed.} The velocity dependent Lyapunov exponent is then
\beq
\lambda(v)=\text{ext}_\nu [j(\nu)-1-(i\nu+d/2-1) v].
\eeq
Setting $\nu=-ir$ as before, this is
\beq
\label{eq:ReggeVDLE}
\lambda(v)=\text{ext}_r [j(-i r)-1-(r+d/2-1) v],
\eeq
i.e. $\lambda(v)+v(d/2-1)$ is the Legendre transform of the function $1-j(-i r)$. We have assumed this to be a concave function of real $r$. It is well known that the Legendre transform takes convex functions to convex functions, which implies that $\lambda(v)$ is concave as a function of real $v$, $\lambda''(v)<0$. It would be interesting to see if this could be proved in general for the VDLE (we know of no counter example).

\subsubsection*{Critical velocity and exchange of dominance}

Now let us explore the consequences of the bound \eqref{eq:Legendreboundisotropic}. Since the Legendre transformation is an involution, $1-j(-i r)$ is also the Legendre transform of $\lambda(v)+v(d/2-1)$. This latter function is just a linear shift of $\lambda(v)$, therefore its Legendre transform is just a shift in the argument of the Legendre transform of $\lambda(v)$ itself. Therefore, \eqref{eq:Legendreboundisotropic} implies that the relation \eqref{eq:ReggeVDLE} can only hold as long as
\beq
\label{eq:jbound}
|j(-i r)-1|\leq 1.
\eeq
If $j(-ir)$ is a convex function, this is guaranteed to break down at some $r$. So what happens then? The answer is the same as many times before: when we deform the integration contour in \eqref{eq:Reggesaddle} to the steepest descent contour, we can cross a pole in $g(\nu)$ which can dominate over the saddle contribution (see Fig.~\ref{fig:exchangeofdom}).  
The simplest possibility for this pole comes from an explicit factor of $\left(\sin \frac{\pi j(\nu)}{2} \right)^{-1}$ \cite{Costa:2012cb}, which picks the $\nu$ where $j=2$, namely the stress tensor, sitting at $\Delta=d$ or $r=i\nu=d/2$. This is precisely the value where \eqref{eq:jbound} gets saturated. It is reasonable to assume that if the stress tensor is the lightest operator on the leading Regge trajectory, there is no non-analyticity in $g(\nu)$ closer than this to the real $\nu$ axis. The contribution of this pole in \eqref{eq:Reggesaddle} gives the exponent
\beq
\label{eq:Rindlerballistic}
\lambda(v)=1-(d-1)v,
\eeq
which saturates the bound \eqref{eq:VDLObound2} with $v^T_B=(d-1)^{-1}$, where the superscript $T$ refers to the stress tensor. This is indeed known to be the butterfly velocity of holographic CFTs on Rindler space \cite{Perlmutter:2016pkf,Ahn:2019rnq}. We will comment more on this butterfly speed towards the end of the section.

The exchange of dominance between the saddle and the pole happens at a critical velocity
\beq
\label{eq:criticalv}
v_*=\frac{1}{i}j'(\nu_T)=\frac{1}{i}j'\le(-i\frac{d}{2}\ri).
\eeq
Let us try to gain some further insight on this $v_*$ by assuming that $j(-ir)$ is a convex function of $r$. In this case, $j$ must be above any of its tangents
\beq
\label{eq:convexbound}
j(-i r_2)\geq j(-ir_1)+[\partial_r j(-i r_1)](r_2-r_1),
\eeq
for any $r_{1,2}$. Take $r_1=d/2$ and $r_2=r_{hs}$, some single trace higher spin operator $O_{hs}$ sitting on the leading large $N$ Regge trajectory with dimension $\Delta_{hs}=d+j(-i r_{hs})-2+\gamma_{hs}$, where $\gamma_{hs}\geq 0$ parametrizes how much $O_{hs}$ is above the unitarity bound $\Delta-j\geq d-2$. Plugging this choice into \eqref{eq:convexbound} (and using \eqref{eq:criticalv}) leads to
\es{vsIneq}{
\frac{j(-i r_{hs})-2}{j(-i r_{hs})-2+\gamma_{hs}}\geq v_*
}
We get the best bound on $v_*$ by minimizing the left hand side of \eqref{vsIneq} over all  $O_{hs}$ sitting on the leading large $N$ Regge trajectory. Unitarity implies $v_* \leq 1$. This is the same inequality as $v_*\leq v^T_B$ in $d=2$, but weaker in higher dimensions. Also, notice that we have just shown in $d=2$ that if there are single trace higher spin operators that are not conserved, then $v_*$ is strictly smaller than $v_B=1$, implying a region in the butterfly cone where chaos is maximal.\footnote{We remind the reader that in $d=2$ the Rindler and the thermal OTOC are the same.}\footnote{Since the Sommerfeld-Watson resummation relies on expansion in terms of global partial waves, the true leading Regge trajectory in $d=2$ just collects the Virasoro descendants of the identity operator (and possibly other operators in a larger chiral algebra) and one might wonder why can we infer anything useful from it. The answer is that the Lyapunov region is controled by the large $N$ Regge trajectory from which all the multi-traces of $T$ are banned. In other words, at large $N$ the Virasoro and global conformal blocks are the same for light external operators.} For example, if there is a spin four single trace with $\gamma_{hs}>0$ we can explicitly bound
\beq
v_* \leq \frac{1}{1+\gamma_{hs}/2}.
\eeq

Assuming convexity, we can also put a \textit{lower bound} on $v^*$ in terms of the Regge intercept, by taking $r_1=d/2$, $r_2=0$ in \eqref{eq:convexbound}. This leads to
\beq
2\frac{1-\lambda_L}{d}\leq v_*,
\eeq
where we have written the result in terms of the hyperbolic space Lyapunov exponent $\lambda_L=j(0)-1$.
Smaller critical velocities therefore require the conventional Lyapunov exponent to be larger. In particular, in case there exists a region with maximal chaos inside the butterfly cone, we have $v_*< v_B=(d-1)^{-1}$ and the above inequality implies that $\lambda_L$ cannot be smaller than $\frac{1}{2}\frac{d-2}{d-1}$. In case the theory has a weak coupling limit, this suggests that the region with maximal chaos disappears at a finite coupling, since we must have $v_*\rightarrow v_B$ before the Lyapunov exponent could reach zero. 
In fact, at zero coupling, the $j>1$ part of the Regge trajectory is just given by the unitarity bound $j=\Delta-d+2$, so $v_*=1$, which is much larger than $(d-1)^{-1}$. As we turn on the coupling, we expect small corrections to this. 

\subsubsection*{Critical velocity in planar $\mathcal{N}=4$ SYM}

A concrete example is the $\mathcal{N}=4$ SYM theory in $d=4$ in the planar, $N\to \infty$ limit. The leading Regge trajectory at weak coupling was analyzed in \cite{Costa:2012cb} based on results of \cite{Kotikov:2002ab}, and to leading order, the critical velocity is:\footnote{The Regge trajectory at weak coupling in a neighborhood of the stress tensor, $r=i\nu=2$ is given by $j(r)=r-{\lambda_\text{'t Hooft}\, H(r-2)/(2\pi^2)}+O\le(\lambda_\text{'t Hooft}^2\ri)$, where $H(n)$ is the $n$th harmonic number.}
\beq
v_*=\frac{1}{i} j' \left( -2 i \right) =1-\frac{\lambda_\text{'t Hooft}}{12} + \cdots\,,
\eeq
where $\lambda_\text{'t Hooft} =g_{YM}^2 N$ is the 't Hooft coupling. Therefore, one cannot see the region of maximal chaos in perturbation theory. At strong coupling one has $v_*=2/\sqrt{\lambda_\text{'t Hooft}}$ \cite{Brower:2006ea,Cornalba:2007fs,Costa:2012cb}, which is small as expected.\footnote{The Regge trajectory at strong coupling is $j(\nu)=2-{(4+\nu^2)/( 2\sqrt{\lambda_\text{'t Hooft}})}+O\le(1/ \lam_\text{'t Hooft}\ri)$. } 

Remarkably, one can determine the exact relation between $v_*$ and $\lambda_\text{'t Hooft}$.\footnote{We thank Shota Komatsu for pointing this out to us.} The inverse function of the leading Regge trajectory has an expansion around $j=2$ \cite{Brower:2014wha}
\beq
\Delta(j)=4+\alpha_1(\lam_\text{'t Hooft})(j-2) + \cdots\,,
\eeq
where $\alpha_1$ is the exact slope function of Basso \cite{Basso:2011rs,Gromov:2012eg}:
\beq
\alpha_1(\lambda)= \frac{\sqrt{\lambda}}{2} Y_2(\sqrt{\lambda})\,,
\eeq
with $Y_n(x)=I_n'(x)/I_n(x)$ and $I_n(x)$ the $n$-th modified Bessel function. This leads to the critical velocity $v_*=1/\alpha_1(\lam_\text{'t Hooft})$. This expression is consistent with both the weak and the strong coupling expansions above. The critical and butterfly velocities agree when $v_*=1/3$, or when $\lam_\text{'t Hooft}\approx 37.7384$. Above this coupling, the edge of the butterfly front is ballistic with maximal Lyapunov exponent.

\begin{figure}[!h]
\begin{center}
\includegraphics[width=0.7\textwidth]{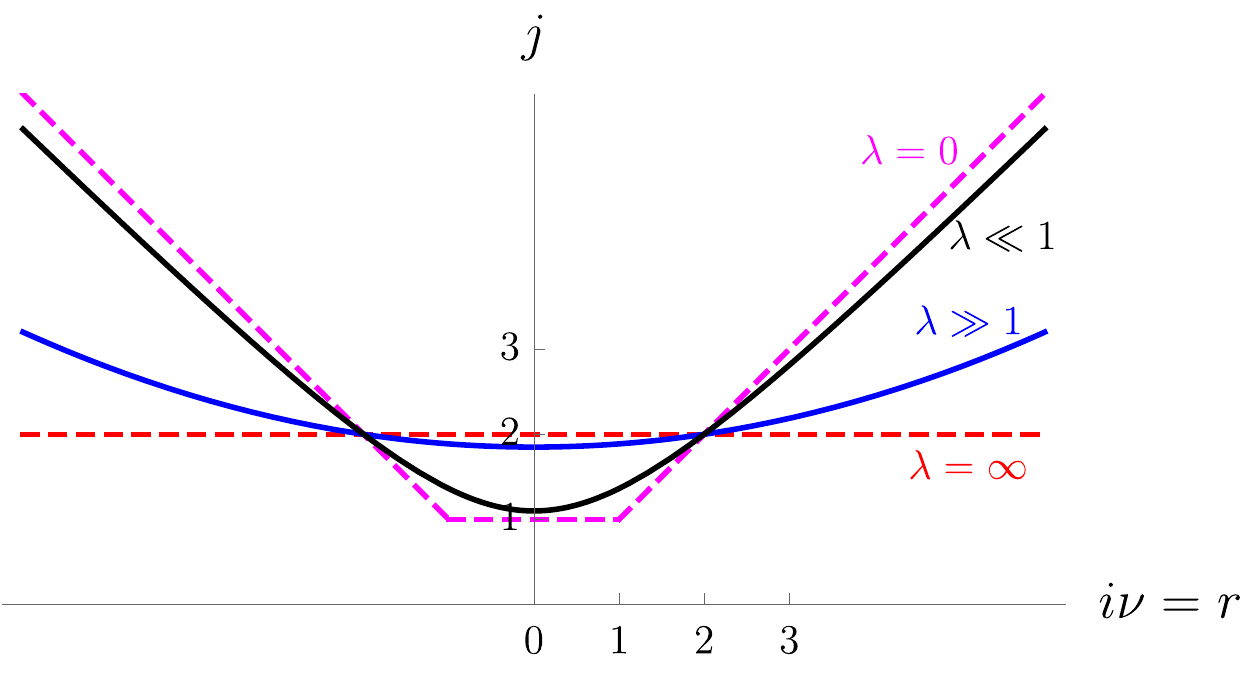}
\caption{Cartoon of the 't Hooft coupling $(\lambda_\text{'t Hooft})$ dependence of the leading large $N$ Regge trajectory in $\mathcal{N}=4$ SYM. The derivative at the stress tensor gives the critical velocity, which is $v_*=1$ at zero coupling, and smoothly goes over to $v_*=0$ at infinite coupling. The region of maximal chaos appears when $v_*<1/3$. \label{fig:reggecartoon}}
\end{center}
\end{figure}

A cartoon of $j(\nu)$ along the imaginary axis is shown on Fig.~\ref{fig:reggecartoon}, adapted from \cite{Costa:2012cb}. There is a lot more known about the leading large $N$ Regge trajectory in this model, for the most recent progress and references, see \cite{Alfimov:2018cms}.

\subsubsection*{Comparison with the light cone OPE}

The light cone limit $\bar z \rightarrow 1$, $z=\text{fixed}$, organizes the OPE according to twist $(\Delta-J)/2$. Assuming that the stress tensor is the only single trace operator with minimal twist, the leading contribution to the four point function, written in our kinematic variables \eqref{eq:cross1}, is \cite{Hartman:2015lfa,Hartman:2016lgu,Perlmutter:2016pkf}
\beq
\label{eq:lightconeOPE}
f(z,\bar z) \sim 1+ \# U^{1-d} e^T [1+O(Ue^{-T}) + \cdots] +\cdots,
\eeq
where we take $1-z=Ue^{-T}$ fixed, but small as we send $T\rightarrow \infty$. In terms of our scaling $U=e^{vT}U_0$, this requires $v=1$ and $U_0$ small. On the other hand, we have essentially shown that in large $N$ CFTs, this form remains valid whenever $v\geq v_*$. Strictly speaking, we have only shown this when the contribution \textit{grows}, that is, when $v_*\leq v\leq (d-1)^{-1}$, and also we have not explored the coefficient of the growing piece. Nevertheless, it seems to be a reasonable possibility that the light cone OPE remains valid in the regime $v_*\leq v \leq 1$. It would be interesting to explore this further.

\subsubsection*{Butterfly speed on Rindler space}

When $v_*< v^T_B=(d-1)^{-1}$, the exchange of dominance between the saddle and the pole in \eqref{eq:Reggesaddle} happens in the region where $\lambda(v)>0$. If this is the case, the VDLE is given by \eqref{eq:Rindlerballistic} for $v>v_*$ and the true butterfly speed is $v_B=v^T_B=(d-1)^{-1}$.\footnote{Note that this butterfly velocity can be formally inferred from the light cone OPE \eqref{eq:lightconeOPE}, as done in \cite{Perlmutter:2016pkf}. However, as mentioned before, the light cone OPE is known to converge only when $v=1$, so the true butterfly velocity can be outside of its regime of validity.} We expect this to be the case for all sufficiently strongly coupled large $N$ CFTs, since stringy corrections in holography suggest that $v_*$ at strong coupling is proportional to the inverse coupling.\footnote{For instance, in holography, higher derivative corrections to $v_B$ on Rindler-AdS should be forbidden. This sounds reasonable, since shockwaves in AdS do not receive $\alpha'$ corrections \cite{Horowitz:1999gf} (we thank Viktor Jahnke for pointing this out to us).} We argued that in $d>2$ as we decrease the coupling, before the Lyapunov exponent could become too small, $v_*$ must reach $v_B^T$ and then pass it. Since at $v_*$ both the value and the first derivative of the saddle and the pole contribution to $\lambda(v)$ must agree, assuming concavity of the saddle contribution to $\lambda(v)$, we see that when $v_*>(d-1)^{-1}$, $\lambda(v)$ must cross zero at some $v_B < (d-1)^{-1}$. Therefore, any large $N$ CFT must have butterfly speed on Rindler space that is at most as big as the holographic result, that is
\beq
v_B \leq \frac{1}{d-1}.
\eeq
In $d=2$ this is an equality and $v_B=1$.

\section{Rotating ensembles}\label{sec:Rotating}

\subsection{General remarks}

The generalization of the chaos bound of \cite{Maldacena:2015waa} to ensembles involving chemical potentials for other spacetime generators than the Hamiltonian is rather straightforward. We simply replace in \eqref{eq:OTOCdef}
\beq
y \rightarrow \frac{e^{-\frac{\theta_a}{4} Q_a}}{Z^{1/4}}, \quad Z=\text{Tr} e^{-\theta_a Q_a},
\eeq
where $\theta_a$ are chemical potentials for the generators $Q_a$. In order for the trace to converge in \eqref{eq:OTOCdef}, we need to choose the $Q_a$ and the $\theta_a$ such that $\theta_a Q_a$ is a positive operator. In this case, we can just think of this operator as a Hamiltonian and immediately deduce the bound\footnote{This bound applies to OTOCs which are symmetrically regulated with the positive operator $\theta_a Q_a$. For possible issues with the regulator dependence of the chaos bound, see \cite{Liao:2018uxa,Romero-Bermudez:2019vej}.}
\beq
\label{eq:trivialbound}
\frac{|\theta_a [Q_a]^\mu\partial_{\mu}f|}{1-f} \leq 2\pi,
\eeq
where we assumed the operators $V,W$ to be scalars for simplicity, on which the generators $Q_a$ act as
\beq
[Q_a]^\mu\partial_{\mu} V =i [Q_a,V].
\eeq
This equation defines the coefficients $[Q_a]^\mu$.

The simplest example of this is when we take the ensemble to be just a boosted thermal ensemble,  $\theta_a Q_a=\beta^\mu P_\mu$, with $\beta^\mu$ timelike. If the velocity corresponding to the boost is ${\bf v}$ then
\es{Betamu}{
\beta^\mu=  \frac{\beta}{\sqrt{1-{\bf v}^2}}(1,{\bf v})\,, \qquad \beta=\sqrt{-\beta^\mu \beta_\mu}\,.
}
In a relativistic theory this ensemble is of course equivalent with the unboosted thermal ensemble. The bound then reads as
\beq
\frac{|\beta^\mu\partial_{\mu}f|}{1-f} \leq 2\pi.
\eeq
It is interesting to ask if we can bound $|\partial_t f|$ alone when the ensemble is boosted. For a Lorentz invariant theory, the formula \eqref{eq:OTOCVDLE} defining the VDLE, can be translated to the form \eqref{eq:OTOCdecay} for the boosted ensemble, when we do not scale ${\bf x}$ with $t$:
\beq
f_{\beta^\mu}(t)\approx 1-\epsilon\, e^{\lambda_L(\beta^\mu)\, t} + \cdots.
\eeq
The relation of this ``boosted" Lyapunov exponent to the VDLE is
\beq
\lambda_L(\beta^\mu)= \frac{1}{\sqrt{1-{\bf v}^2}} \lambda( {\bf v})\,, 
\eeq
where the $ 1/\sqrt{1-{\bf v}^2}$ factor comes from time dilation. The VDLE bound \eqref{eq:VDLObound2} then reads as
\beq
|\lambda_L(\beta^\mu)|\leq \frac{2\pi}{\beta}\, \frac{1-\frac{|v|}{v_B}}{\sqrt{1-v^2}} \,.
\eeq
In light cone coordinates along the direction of the boost, one may write the components of the boosted temperature as
\beq
\beta_+ = \sqrt{\frac{1+v}{1-v}}\beta\,, \quad \beta_- = \sqrt{\frac{1-v}{1+v}}\beta\,,
\eeq
where $v$ carries sign here. One can then give a simple bound on $\lambda_L$ using these chiral temperatures and that $v_B\leq 1$ in a Lorentz invariant theory:
\bea
\label{eq:2dboostedbound}
|\lambda_L(\beta_+,\beta_-)|&\leq \frac{2\pi}{\beta} \sqrt{\frac{1-|v|}{1+|v|}}\\
&=\min \left\lbrace \frac{2\pi}{\beta_+}, \frac{2\pi}{\beta_-} \right\rbrace.
\eea

Another typical ensemble to consider is obtained by putting a theory on the sphere and turning on an angular velocity (chemical potential). In this case, $\theta_a Q_a=\beta (H + \omega J)$, where $J$ is a component of the angular momentum operator. We can choose $\omega$ such that this is guaranteed to be positive. If the theory is a CFT, this is ensured by unitarity bounds. In this case, the bound \eqref{eq:trivialbound} applies to a combination of a time derivative and an infinitesimal rotation.

We want to highlight a notational issue that could  potentially be confusing. In analogy with the rotating case, it sounds reasonable to write in the boosted case
$\theta_a Q_a=\beta'\le(H+{\bf v}\cdot {\bf P}\ri)$. But in the notation of \eqref{Betamu}, $\beta'\neq \beta$, instead $\beta'$ is the temporal component of $\beta^\mu$. For future reference, we note that the chiral temperatures are related to this as
\es{betapm}{
\beta_\pm=(1\pm v)\beta'\,.
}

\subsubsection*{Comments on the vacuum block approximation}

Before moving on, let us briefly explain how the bound \eqref{eq:2dboostedbound} is obeyed in a 2d CFT in a vacuum Virasoro block approximation \cite{Roberts:2014ifa}, since there is a small subtlety that has lead to some confusion in the literature \cite{Poojary:2018esz,Jahnke:2019gxr,Halder:2019ric}. In a 2d CFT in a thermal state on the line, one can map the four point function to flat space using the exponential map, and therefore it only depends on two cross ratios $z$ and $\bar z$ (this is a special case of the discussion in sec. \ref{sec:Regge}.) The OTOC is obtained by sending one cross ratio to the second sheet $ z\rightarrow e^{2\pi i} z$, while leaving the other $\bar z$ untouched. In the vacuum block approximation, one approximates the four point function as 
\beq
\label{eq:vacuumblock}
f\approx \mathcal{F}(1-z) \mathcal{F}(1-\bar z),
\eeq
where $\mathcal{F}(z)$ is the Virasoro vacuum block in the cross channel that is not invariant under the operation $ z\rightarrow e^{2\pi i} z$. Note that in the original unboosted setup of \cite{Roberts:2014ifa}, $z$ depends only on $x-t$ and $\bar z$ depends only on $x+t$, so naively doing the continuation in $z$ would lead to an OTOC that goes like $f\sim 1-\# e^{\frac{2\pi}{\beta}(t-x)}/c$, which contradicts the manifest parity invariance $x\rightarrow -x$ of the correlator. The resolution is the following. The exact four point function must be single valued on the Euclidean sheet $z^*=\bar z$ so it must be invariant under the joint operation $ z\rightarrow e^{2\pi i} z$ and $ \bar z\rightarrow e^{-2\pi i} \bar  z$. However the factorized form \eqref{eq:vacuumblock} is not invariant under this operation. This is because $\mathcal{F}(1-e^{2\pi i}z) \mathcal{F}(1-e^{-2\pi i} \bar z)$ corresponds to the vacuum block in a different OPE channel; see \cite{Maloney:2016kee} for an explanation of the possible channels for the vacuum block. According to the general rule of this approximation, one must take the block in the channel where it is the largest \cite{Asplund:2014coa}. This means that we need to compare the continuations $z\rightarrow e^{2\pi i} z$ and $\bar z\rightarrow e^{2\pi i} \bar z$ in \eqref{eq:vacuumblock} and take the result that is larger in magnitude. This procedure correctly gives $f\sim 1-\# e^{\frac{2\pi}{\beta}(t-|x|)}/c$ in the unboosted case, consistently with the $x\rightarrow -x$ symmetry. It also gives the Lyapunov exponent $\lambda_L= \min \lbrace \frac{2\pi}{\beta_-},\frac{2\pi}{\beta_+} \rbrace$ in the boosted case, saturating our bound \eqref{eq:2dboostedbound}.

\subsection{OTOC in a rotating black hole}

Below we present the computation of the OTOC in a rotating thermal ensamble in a 2d CFT  with an Einstein gravity dual. Most of the computations were done in \cite{Fu:2018oaq,Poojary:2018esz,Jahnke:2019gxr}; the novelty of this section is a careful interpretation of the results and their comparison with other results in this paper.

\subsubsection*{Shockwaves in rotating BTZ}

The metric of the rotating BTZ black hole is
\es{RotatingBTZ}{
ds^2&=-g(r) dt^2+ {dr^2\ov g(r)}+r^2\le(d\varphi-{r_+r_-\ov \ell r^2}dt\ri)^2\,,\\
g(r)&={(r^2-r^2_+)(r^2-r^2_-)\ov \ell^2 r^2}\,,
}
where $\varphi\sim \varphi+2\pi$ is an angular coordinate and $\ell$ is the AdS radius. It will be convenient for us to think of the boundary theory as living on an a circle of radius $\ell$.\footnote{The boundary metric \eqref{RotatingBTZ} on the cutoff surface $r=r_c$ is of the form $r_c^2/\ell^2 (-dt^2+\ell^2 d\varphi^2)$, so in the conformal frame where we drop the prefactor, the boundary circle has length $2\pi \ell$.} The inverse temperature and angular velocity  of the boundary theory is
\es{TempChem}{
\beta={2\pi \ell^2 r_+\ov r_+^2-r_-^2}\,, \qquad \om={r_-\ov \ell r_+}\,.
}
This black hole only dominates the canonical ensemble for ${\beta/ \ell}<{2\pi/\sqrt{1-(\om \ell)^2}}$,  at temperatures above the Hawking-Page transition \cite{Hawking:1982dh}. Below the transition (for large $\beta/ \ell$) the dominant saddle is thermal AdS and the BTZ black hole is a subdominant saddle. The results discussed below should be interpreted with this subtlety in mind. 

The OTOC for $t\gg \beta$ is given by 
\es{OTOC}{
f(t,\varphi)=1-{\#\ov N^2}e^{{2\pi\ov \beta}t}h\le({\om t}-\varphi\ri)+O\le(1\ov N^4\ri)\,,
}
where $h(\phi)$ is the shock wave profile on the black hole horizon in co-rotating coordinates $\phi\equiv \varphi-{\om t}$. Note that it is $h(-\phi)=h\le({\om t}-\varphi\ri)$ that appears in the expression \eqref{OTOC}, for a reminder of why this is we refer to  \cite{Jahnke:2019gxr}. $h(\phi)$ solves the equation \cite{Fu:2018oaq,Jahnke:2019gxr}:
\es{hEq}{
&h''(\phi)-{2r_-\ov \ell} h'(\phi)-{r_+^2-r_-^2\ov \ell^2}h(\phi)=\#\delta(\phi)\,,
}
and is given by
\es{hEq2}{
&h(\phi)={\exp\le(-{2\pi \ov \hat\beta (1+\hat\om)}\le[\phi  \pmod{2\pi} \ri]\ri)\ov 1-\exp\le(-{4\pi^2  \ov \hat \beta (1+\hat \om)}\ri)}+{\exp\le({2\pi  \ov \hat \beta (1-\hat\om)}\le[\phi  \pmod{2\pi} \ri]\ri)\ov \exp\le({4\pi^2  \ov \hat \beta (1-\hat \om)}\ri)-1}\,,\\
&\hat \beta\equiv{\beta\ov \ell}\,, \qquad \hat \om \equiv \ell \om\,, \qquad \hat \om\in(-1,1)\,,
}
where we used \eqref{TempChem} to convert to field theory quantities.
It is {\it absolutely crucial} that we take into account the $\pmod{2\pi}$ in this formula in the following; this is the only improvement we have over the analysis of \cite{Poojary:2018esz,Jahnke:2019gxr}. Forgetting the periodicity of $h(\phi)$ can lead to erroneous conclusions about the growth of the OTOC.

\subsubsection*{Analysis of the OTOC and the instantaneous Lyapunov exponent}

\begin{figure}[!h]
\begin{center}
\includegraphics[scale=0.6]{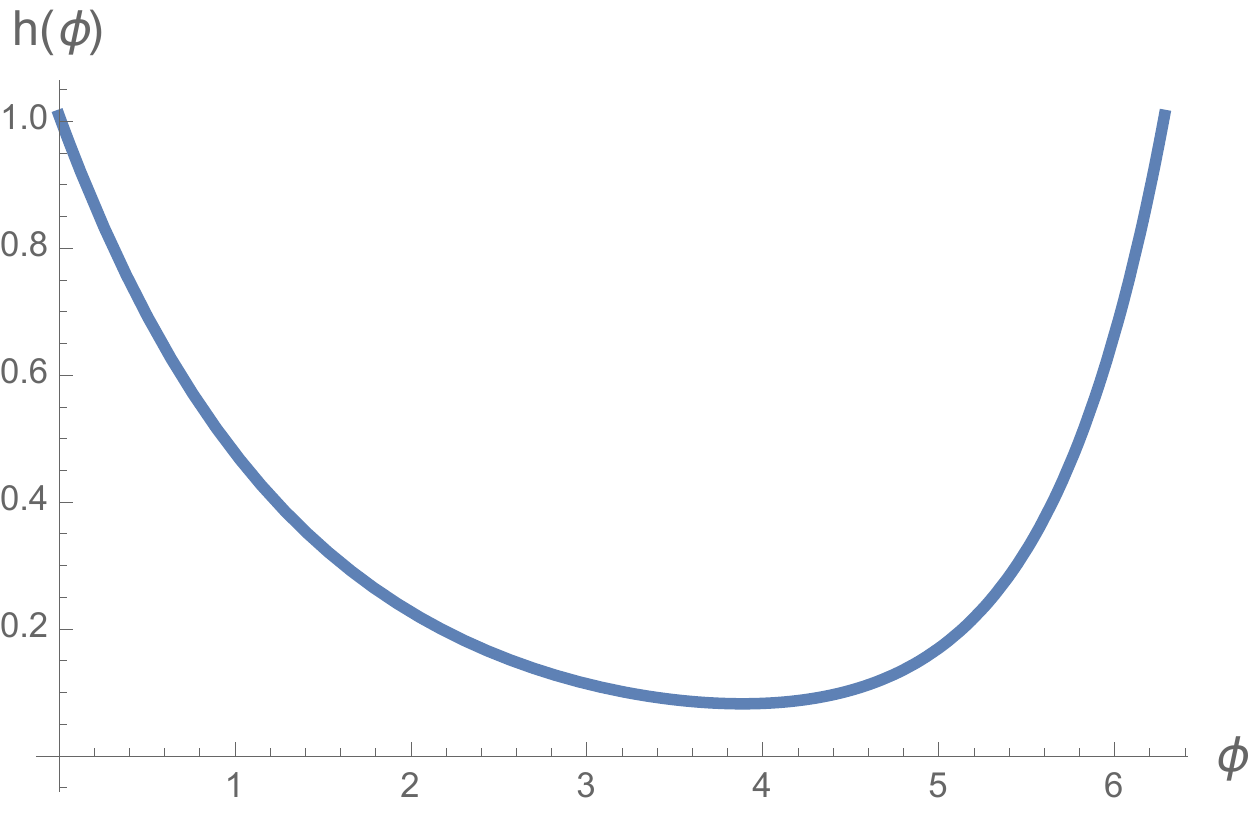}
\caption{$h(\phi)$ for $\hat\beta=2\pi,\, \hat \om=1/3$. \label{fig:fplot}}
\end{center}
\end{figure}

\begin{figure}[!h]
\begin{center}
\includegraphics[scale=0.6]{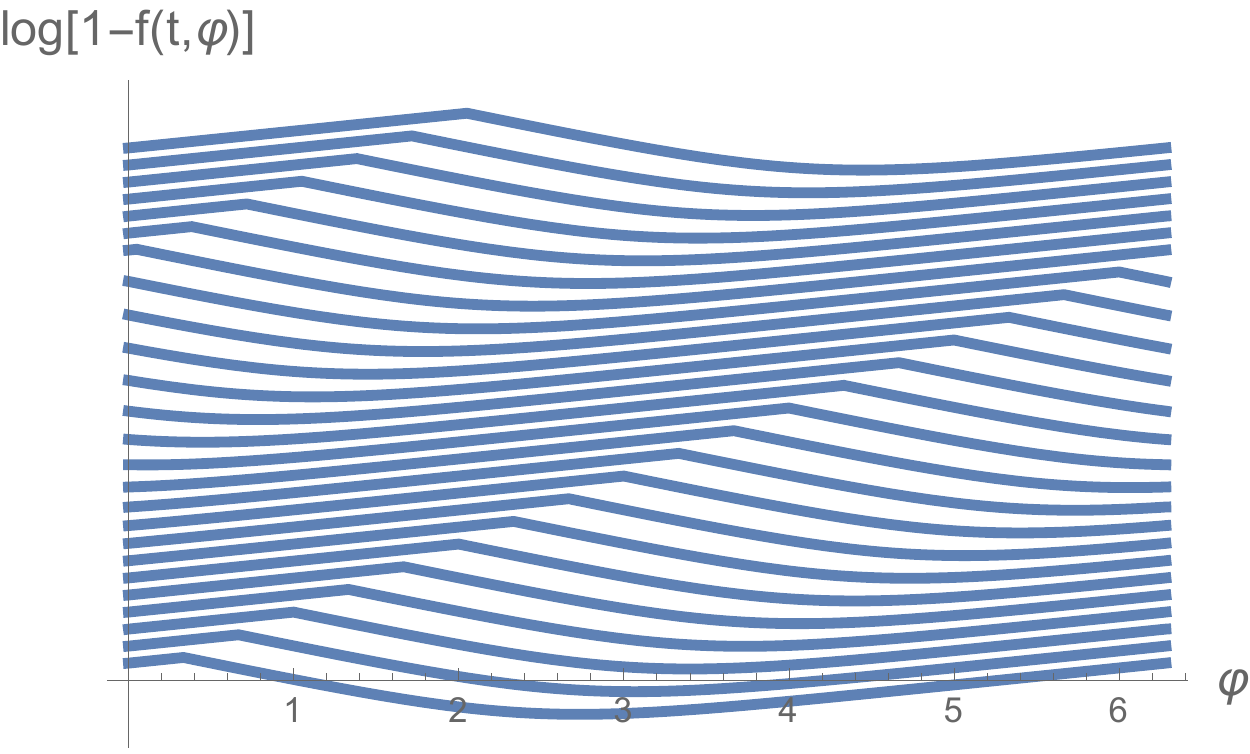}\hspace{1cm}
\includegraphics[scale=0.3]{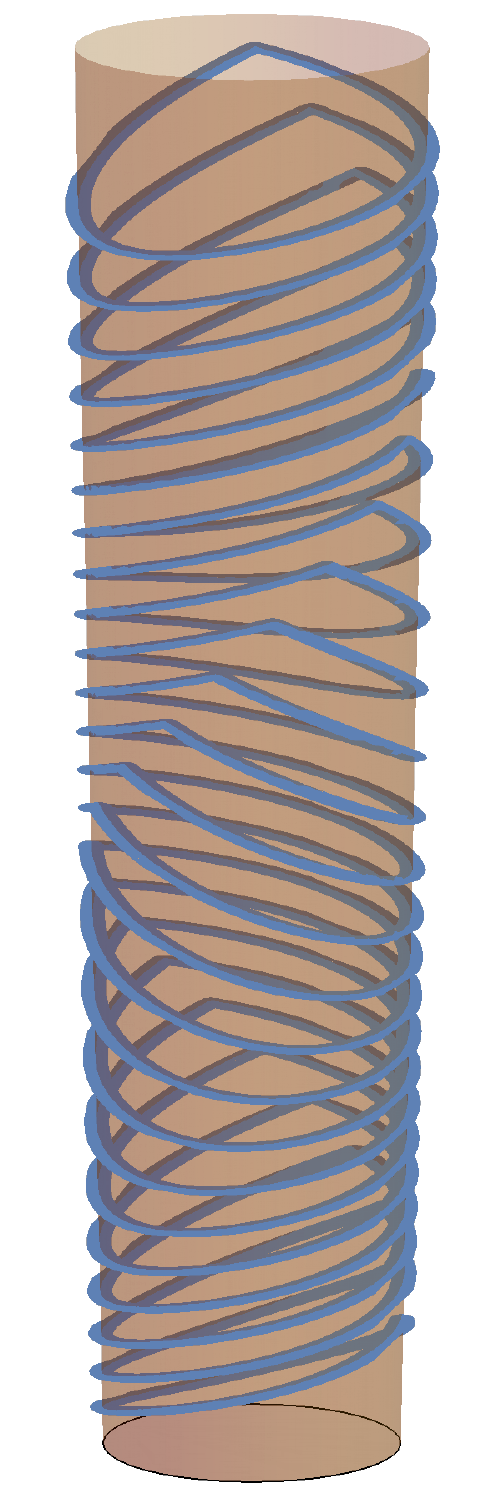}
\caption{Snapshots of $\log \le[1-f\le(t,\varphi\ri)\ri]$  for $\hat\beta=2\pi,\, \hat \om=1/3$ at different times as a function of $\varphi$. On the right we plot the same snapshots on the cylinder so that the periodicity of $\varphi$ is manifest. \label{fig:snapshots}}
\end{center}
\end{figure}

In Fig.~\ref{fig:fplot} we plot $h(\phi)$, this plot is all we need to understand the other figures. In Fig.~\ref{fig:snapshots} we show $ f(t,\varphi)$ as a function of $\varphi$ at different times. The cusp in the function originates from the $\delta$-function in \eqref{hEq}. 


\begin{figure}[!h]
\begin{center}
\includegraphics[scale=0.6]{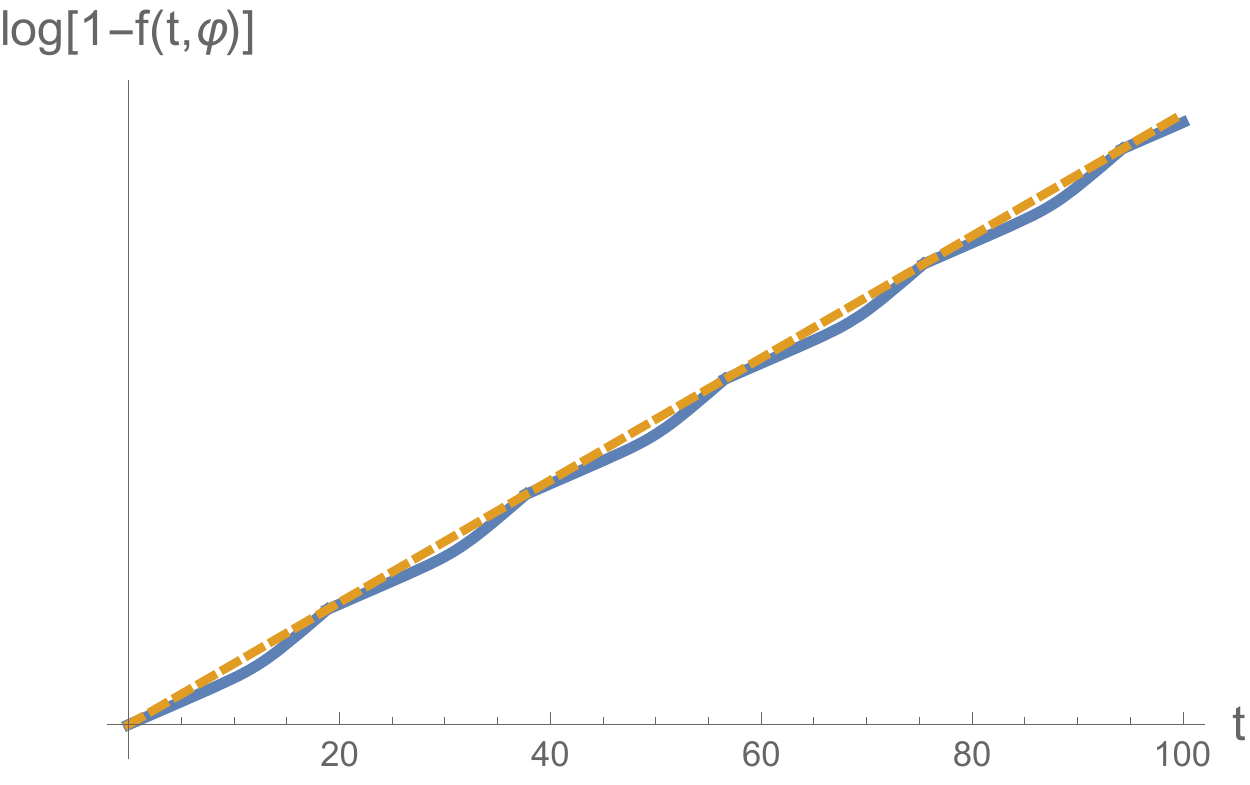}
\caption{The blue line is $\log \le(1-f\le(t,\varphi\ri)\ri)$ at $\varphi=0$ as function of time for $\hat\beta=2\pi,\, \hat \om=1/3,\, \ell=1$. The orange dashed line is (the logarithm of) $e^{{2\pi\ov \beta}\, t}$. The blue curve is a periodic modulation on this exponentially growing piece.\label{fig:timedep}}
\end{center}
\end{figure}

In  Fig.~\ref{fig:timedep} we fix $\varphi$ and plot the OTOC as a function of time: we see that the function grows exponentially as $e^{{2\pi\ov \beta}t}$ with a periodic modulation. This modulation is easy to understand, 
$h(\phi)$ is a periodic function of $\phi$, hence $ f(t,\varphi)$ is a periodic function of both $t$ and $\varphi$, with period ${2\pi \ov \om}$ and $2\pi$ respectively. Since, the OTOC on average decreases with exponent ${2\pi\ov \beta}$ we can say that the average Lyapunov exponent is:
\es{Lyap}{
\overline{\lam_L}={2\pi\ov \beta}\,.
}

\begin{figure}[!h]
\begin{center}
\includegraphics[scale=0.6]{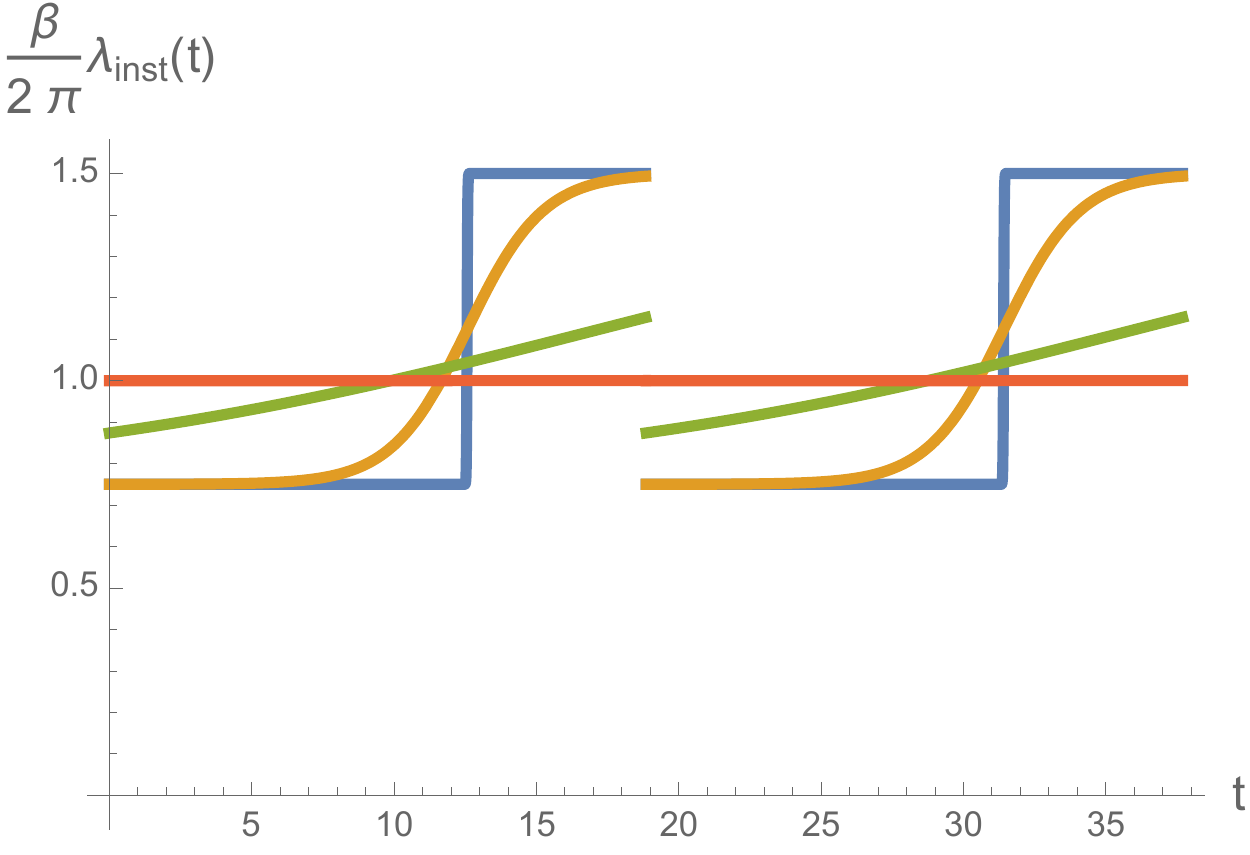}
\caption{The instantaneous Lyapunov exponent $\lam_\text{inst}(t)$ as a function of time. We set $\hat \om=1/3,\, \ell=1$ and plot ${\beta\ov 2\pi}\,\lam_\text{inst}(t)$ for $\beta\to0$ (blue), $2\pi$ (orange), $16\pi$ (green), and $\infty$ (red). 
\label{fig:LyapInst}}
\end{center}
\end{figure}

%\pagebreak
We can also define a time dependent, instantaneous exponent by\footnote{Including $\varphi$ in the expression below would shift $\lam_\text{inst}(t)$ in time, but would not change its features.}
\es{LyapInst}{
\lam_\text{inst}(t)\equiv {\abs{\p_t  f(t,0)}\ov 1-  f(t,0)}\,, \qquad  \beta\ll t \ll \beta\log N\,,
}
which has the explicit expression
\es{LyapInst2}{
\lam_\text{inst}(t)={2\pi\ov \beta}+{\om h'(\om t)\ov h(\om t)}\,.
}
We plot some examples in Fig.~\ref{fig:LyapInst}. Some notable special cases are worth understanding. For $\om=0$ or in the small temperature limit, ${ \beta\ov \ell}\to \infty$, $\lam_\text{inst}(t)={2\pi\ov \beta}$. In the high temperature limit, ${ \beta\ov \ell}\to 0$, $\lam_\text{inst}(t)$ is a step function jumping between the values ${2\pi\ov (1+\abs{\hat \om})\beta}$ and ${2\pi\ov (1-\abs{\hat \om})\beta}$. The jump happens at $t={1+\abs{\hat\om}\ov \abs{\hat\om}}\pi \ell$. The time of the jump is proportional to the system size and for times much smaller than this, $\beta\ll t\ll \ell$,  $\lam_\text{inst}(t)$ saturates the bound derived for boosted thermal systems on the line, \eqref{eq:2dboostedbound}, as expected. To see this, we have to remind ourselves that in the language of the discussion around \eqref{betapm}, the $\beta$ used here is really $\beta'$ and  $\hat\om=v$, hence the values that the step function takes are ${2\pi\ov \beta_\pm}$. In previous work these extremal values were found, and referred to as $\lam_\pm$ \cite{Poojary:2018esz,Jahnke:2019gxr}.

It is also interesting to implement the decompactification limit $\ell\to\infty$ on the OTOC. We introduce $x\equiv \ell \varphi$ that we keep finite as we take $\ell\to\infty$. Using also that $\phi={x-\hat\om t\ov \ell}\ll 1$, the OTOC simplifies to:
\es{OTOCLine}{
 f(t,x)=1-{\#\ov N^2}\begin{cases}
\exp\le({2\pi \ov \beta (1+\hat\om)}(t+x)\ri) \qquad & (x\leq\hat \om t)\,,\\
\exp\le({2\pi \ov \beta (1-\hat\om)}(t-x)\ri) \qquad & (x>\hat \om t)\,.
\end{cases}
}
The Lyapunov exponent is $\lam_L={2\pi \ov \beta (1+\abs{\hat\om})}\leq {2\pi \ov \beta }$, which does not exceed the maximal known value, $ {2\pi \ov \beta }$. Instead, it saturates the stronger bound \eqref{eq:2dboostedbound} valid for boosted ensembles. The speed dependent Lyapunov exponent is
\es{Lam(v)}{
\lam(v)={2\pi \ov \beta}\begin{cases}
{1+v\ov 1+\hat\om} \qquad & (v\leq\hat \om )\,,\\
{1-v\ov 1-\hat\om} \qquad & (v>\hat \om )\,,
\end{cases}
}
which again cannot exceed ${2\pi \ov \beta }$, and is only equal to it when $v=\hat\om$.\footnote{It is interesting to consider the extremal case $\hat \om=-1$, where we get $\lam(v)={2\pi \ov \beta}\,{1-v\ov 2}$ for $-1<v<1$. It was noticed by Zhenbin Yang that this is the shifted version of VDLE in the chiral SYK model in the limit of maximal interaction strength $\mathcal{J}=1$: $\lam(v)={2\pi \ov \beta}\le(1-v/ 2\ri)$ for $0<v<2$. This observation deserves better understanding.}

Note that for $\hat\om =0$, the expression \eqref{OTOCLine} can be written in the form $\exp\le({2\pi \ov \beta }(t-\abs{x})\ri)$, which is familiar from the work of Shenker and Stanford. For $\hat\om\neq 0$ \eqref{OTOCLine} is simply a boost of this expression, as the ``rotating'' grand canonical ensemble is simply a boosted canonical ensemble for $\ell\to\infty$. Concretely, in a reference frame boosted by $-\hat \om$:
\es{tmAbsx}{
t-\abs{x}&=\begin{cases}
\sqrt{1-\hat \om\ov 1+\hat \om}\, (t'+x') \qquad & (x'\leq\hat \om t')\,,\\
\sqrt{1+\hat \om\ov 1-\hat \om}\, (t'-x') \qquad & (x'>\hat \om t')\,.
\end{cases}
}
Reminding ourselves that, according to the discussion around \eqref{betapm}, $\beta$ in \eqref{OTOCLine} is really $\beta'=\beta/\sqrt{1-\hat\om^2}$, the above then combine to give \eqref{OTOCLine} (with the primes dropped).

If we want to talk  about the butterfly speed sensibly, we have to restrict attention to $t<\pi\ell$. Combining this with the requirement $t\gg \beta$, we have to be in the ${\beta\ov \ell}\ll 1$ regime of parameters. In that regime the OTOC takes the form \eqref{OTOCLine}, and the butterfly speed is $v_B=1$ irrespective of $\hat\om$.

Finally, it is also easy to check that \eqref{OTOC} satisfies
\beq
\frac{|(\partial_t+\omega \partial_\varphi) f(t,\varphi)|}{1- f(t,\varphi)}=\frac{2\pi}{\beta},
\eeq
that is, it saturates the modified chaos bound \eqref{eq:trivialbound} assoicated to the positive operator $\theta_a Q_a=\beta(H+\omega J)$, generating the rotating ensemble, viewed as a Hamiltonian.

\section*{Acknowledgements}

We thank Viktor Jahnke, Juan Maldacena, Dalimil Mazac, Douglas Stanford, and Brian Swingle for useful discussions, and Viktor Jahnke, Yingfei Gu, Douglas Stanford, Brian Swingle and Zhenbin Yang for comments on the manuscript. MM is supported by the Simons Center for Geometry and Physics. GS acknowledges support from the Simons Foundation through the It from Qubit collaboration (385592, V. Balasubramanian) and was partially supported by FWO-Vlaanderen through project G044016N, and by Vrije Universiteit Brussel through the Strategic Research Program ``High-Energy Physics".

%\clearpage
\bibliographystyle{utphys}
\bibliography{chaosbfcone}

\end{document}